\documentclass[reprint,apl,amsmath,amssymb,doi,superscriptaddress,showkeys]{revtex4-2}

\usepackage{graphicx} 
\usepackage{siunitx}
\usepackage{float}
\usepackage{multirow}

\usepackage{booktabs}
\usepackage{longtable}
\usepackage[colorlinks=true,
    citecolor=black,
    linkcolor=black,
    urlcolor=blue,
    pagebackref=false]{hyperref}
\usepackage{siunitx}
\usepackage{mhchem}

\graphicspath{%
  {./notes-and-scripts/}
  {./../notes-and-scripts/}
  {./figs/}
  {./../figs/}
}

\usepackage{subfiles} %

\newcommand{\subs}[1]{_{\mathrm{#1}}}

\newcommand{\fig}[1]{Figure~\ref{#1}}
\newcommand{\eqn}[1]{Equation~(\ref{#1})}
\newcommand{\eqns}[2]{Equations~(\ref{#1}) and (\ref{#2})}

\begin{document}
\sloppy
\def\createbibliography{} %

\title{Electrical Scanning Probe Microscope Measurements Reveal 
Surprisingly High Dark Conductivity in Y6 and PM6:Y6 and 
Non-Langevin Recombination in PM6:Y6}
\author{Rachael L. Cohn}
\affiliation{Department of Chemistry and Chemical Biology, Cornell University, Ithaca, New York 14853 USA}
\author{Christopher A. Petroff}
\affiliation{Department of Chemistry and Chemical Biology, Cornell University, Ithaca, New York 14853 USA}
\affiliation{Department of Materials Science and Engineering, Cornell University, Ithaca, New York 14853 USA}
\author{Virginia E. McGhee}
\affiliation{Department of Chemistry and Chemical Biology, Cornell University, Ithaca, New York 14853 USA}
\author{John A. Marohn}
\email{jam99@cornell.edu}
\affiliation{Department of Chemistry and Chemical Biology, Cornell University, Ithaca, New York 14853 USA}
\date{February 23, 2024}

\keywords{Non-Langevin Recombination,
    Photo-conductivity,
    Bulkheterojunction Solar Cell,
    Nonfullerene Acceptor,
    Scanning Probe Microscopy}

\begin{abstract}
We used broadband local dielectric spectroscopy (BLDS), an electric force microscopy technique, to make non-contact measurements of conductivity in the dark and under illumination of PM6:Y6 and Y6 prepared on ITO and PEDOT:PSS/ITO.
Over a range of illumination intensities, BLDS spectra were acquired and fit to an impedance model of the tip--sample interaction to obtain a sample resistance and capacitance.
By comparing two descriptions of cantilever friction, an impedance model and a microscopic model, we connected the sample resistance inferred from impedance modeling to a microscopic sample conductivity.
A charge recombination rate was estimated from plots of the conductivity versus light intensity and found to be sub-Langevin.
The dark conductivity was orders of magnitude higher than expected from Fermi-level equilibration of the PM6:Y6 with the substrate, suggesting that dark carriers may be a source of open-circuit voltage loss in PM6:Y6.
\end{abstract}

\maketitle

\section{Introduction}

The power conversion efficiency of donor--acceptor solar cells has been increasing rapidly since the introduction of non-fullerene acceptors \cite{Meredith2020jul}.
Solar cells built from the non-fullerene, small-molecule acceptor Y6, the polymer donor PM6, and related molecules, have shown consistently high efficiency \cite{Yuan2019apr,Perdigon-Toro2020jan,Karki2020oct,Wu2020aug,Wen2021apr,Zhang2021jan}, reaching 19\% power conversion efficiency recently \cite{Zhu2022jun}.
It remains puzzling why the best donor--acceptor blends perform so well.
Charge recombination in the best blends is 10's to 1000's of times slower than predicted by Langevin theory \cite{Deibel2009aug,Ferguson2011nov,Murthy2013sep,Savenije2013oct,Proctor2013dec,Howard2014jun,Burke2015apr,Zhang2017sep,Gasparini2017nov,Menke2018jan,Vollbrecht2019jul,Firdaus2020jan,Yin2020feb,Hosseini2020sep,Zhang2021jan}.
This anomalously slow recombination could be due to an improperly estimated local charge density or mobility \cite{Deibel2009aug}; charge trapping \cite{Ferguson2008jul,Maturova2009may,Ferguson2011nov,Savenije2013oct}; the
inhomogeneous nanoscale structure of the donor/acceptor interface \cite{Proctor2013dec,Burke2015apr}; or a built-in electrostatic potential gradient \cite{Karuthedath2021mar}.
Understanding the anomalous charge recombination can potentially reveal new opportunities for further improving efficiency.
With an eye towards microscopically testing competing theories of non-Langevin recombination \cite{Burke2015apr,Liu2015nov,Benduhn2017apr,Hilczer2010apr}, here we introduce a scanning probe measurement of conductivity in donor--acceptor thin films.

The measurement is an electric force microscopy (EFM) technique, broadband local dielectric spectroscopy (BLDS) \cite{Labardi2016may,Tirmzi2017feb,Tirmzi2019feb,Tirmzi2020jun}.
BLDS was introduced by Labardi et al.\ to probe the frequency-dependent dielectric function of insulating polymers \cite{Labardi2016may}.
It was subsequently applied by Tirmzi and coworkers to examine photo-induced electronic and ionic conductivity in lead-halide perovskite films \cite{Tirmzi2017feb,Tirmzi2019feb,Tirmzi2020jun}.
Tirmzi's work was aided by a new theory for computing electric force microscope signals from the sample's complex electrical impedance \cite{Dwyer2019jun}. 
To more quickly identify better organic solar-cell materials, Menke et al.\ argue that donor--acceptor compounds should be screened using techniques that, in contrast with the widely used photoluminescence quenching measurements, probe charge concentration directly \cite{Menke2018jan}.
Such techniques include microwave conductivity \cite{Savenije2013oct}, optical dielectric constant measurements \cite{Lim2019jan}, phase-kick electric force microscopy \cite{Dwyer2017jun}, and BLDS \cite{Tirmzi2017feb,Tirmzi2019feb,Tirmzi2020jun}.
Here we apply BLDS to study conductivity in an organic photovoltaic material for the first time.
We report studies of the conductivity of PM6:Y6, PM6, and Y6 (chemical structures shown in Fig.~\ref{fig:experimental}a) films in the dark and under illumination.

\createbibliography

\section{Results}

\subsection{Estimating resistance}

\sloppy

\begin{figure}
    \centering
    \includegraphics[width=3.3in]{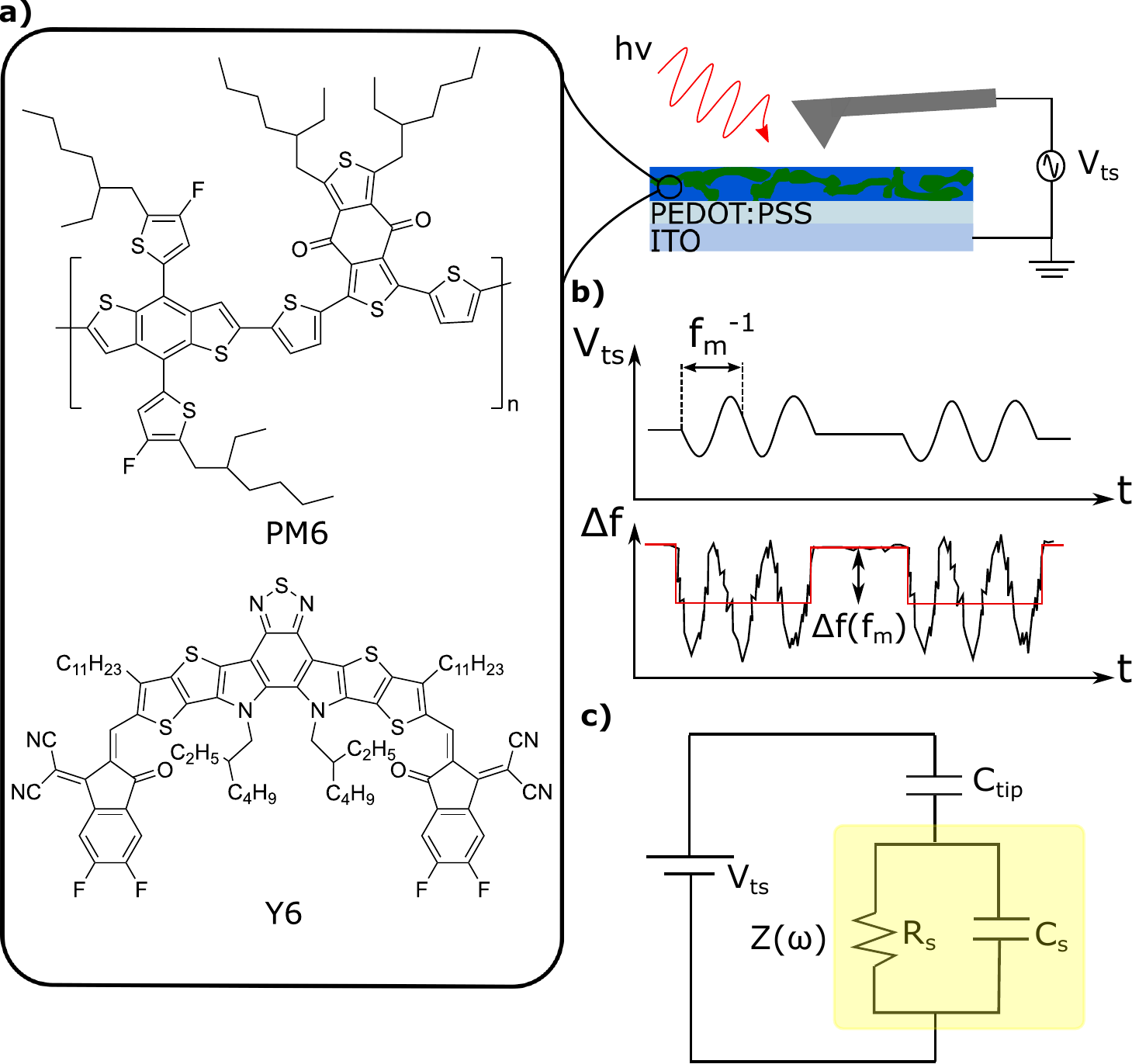}
    \caption{A broadband local dielectric spectroscopy experiment probes photoconductivity in a donor--acceptor solar cell film.
    (a) Experimental setup, showing the structures of the PM6 and Y6 molecules studied here. 
    (b) Applied tip--sample voltage $V_{\mathrm{ts}}$ and the resulting cantilever frequency shift $\Delta f$ versus time $t$; the sinusoidal on--off modulation is depicted as a square wave for simplicity.
    (c) Impedance model of the tip--sample interaction, with $V_{\mathrm{ts}}$ the applied tip--sample voltage, $C\subs{{tip}}$ the tip capacitance, $C\subs{s}$ the sample capacitance, $R\subs{s}$ the sample resistance, and $Z(\omega)$ the sample impedance.}
    \label{fig:experimental}
\end{figure}

\begin{figure*}
    \includegraphics[width=6.6in]{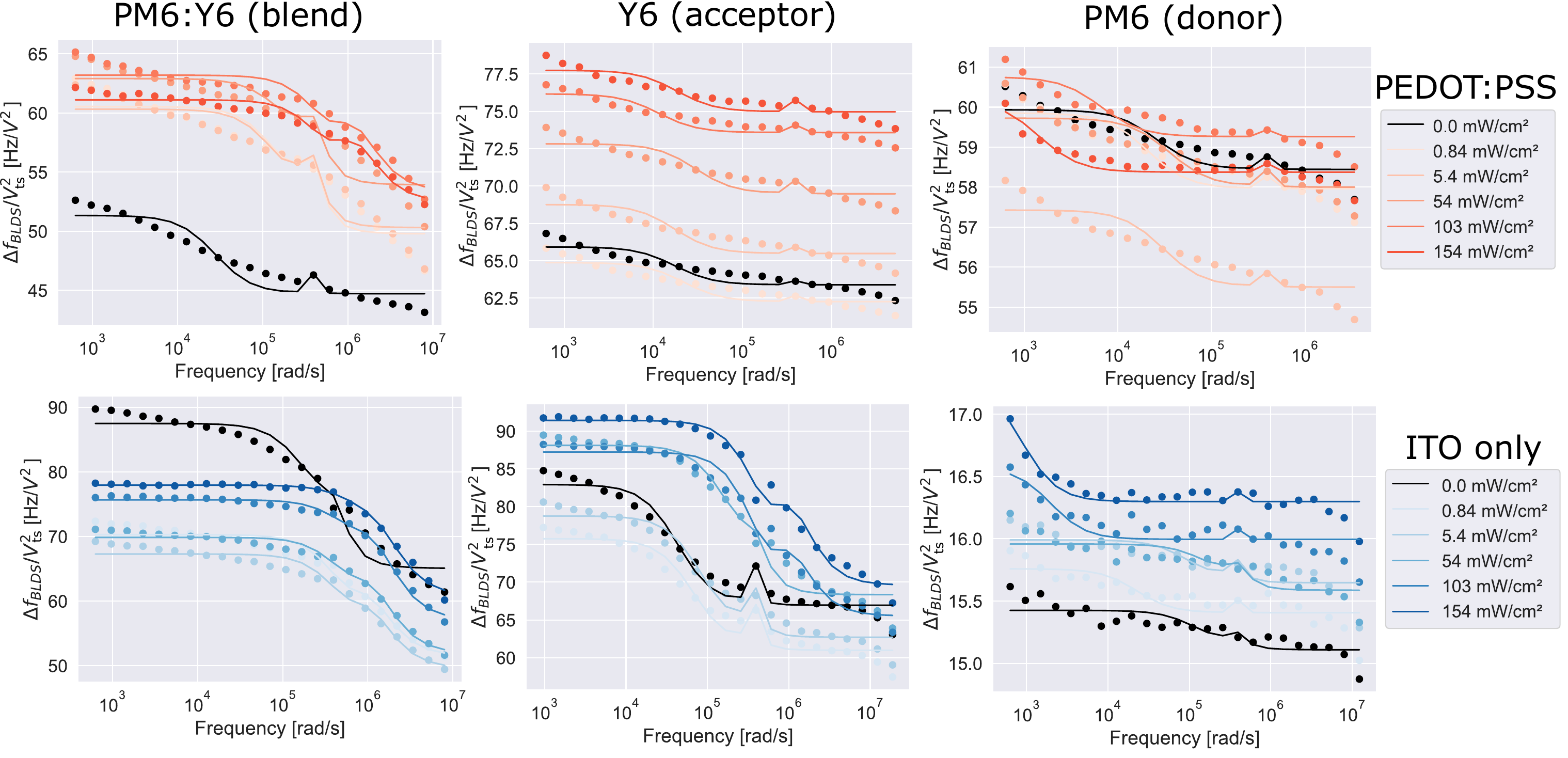}
\caption{Representative broadband local dielectric spectra collected for PM6:Y6 (left), Y6 (middle), and PM6 (right) films with a PEDOT:PSS/ITO contact (top, red) or an ITO only contact (bottom, blue). 
  Films were illuminated from above with a $\lambda = \SI{639.7}{\nano\meter}$ laser at the indicated intensity (right, legend).
  The data were fit to \eqn{eq:f-blds} (lines).}
    \label{fig:blds-data}
\end{figure*}

The BLDS measurement is sketched in \fig{fig:experimental}.
A metal-coated atomic-force microscope cantilever was brought near a sample surface, \fig{fig:experimental}a;
the sample was illuminated from above with light;
and the cantilever's resonance frequency was recorded as a function of time.
A voltage modulation of amplitude $V\subs{ts}$ and oscillation frequency $f\subs{m}$ were applied to the cantilever, shifting the cantilever's resonance frequency by an amount $\Delta f$ on average, \fig{fig:experimental}b.
The voltage was additionally on-off modulated at \SI{20}{\hertz} to allow lock-in detection of $\Delta f$.
The \SI{20}{\hertz} Fourier component of the cantilever frequency, $\Delta f_\mathrm{BLDS}$, is plotted versus voltage-modulation frequency $f\subs{m}$ to yield a BLDS spectrum, \fig{fig:blds-data}.
The spectrum is similar to that of an $RC$ circuit, with the $RC$ time constant light-dependent.
Spectra were collected at various visible light intensities.

BLDS spectra were collected for PM6:Y6, Y6, and PM6 prepared on ITO and PEDOT:PSS/ITO.
Representative data are shown in \fig{fig:blds-data} and all data are shown in Figures~S3--S4.
With the sample modeled as a resistor and capacitor operating in parallel, \fig{fig:experimental}c, these data were fitted to \cite{Tirmzi2017feb,Dwyer2019jun}
\begin{multline}
\frac{\Delta f\subs{{BLDS}}(\omega\subs{m})}{V\subs{ts}^2} = 
-\frac{f\subs{c}}{16 k\subs{c}}
\Bigl(
  C_q^{\prime\prime} +
  \Delta C^{\prime\prime} 
  \: \text{Real} \left[ 
   H(\omega\subs{m} + \omega\subs{c})  \right.
   \\
   \left. + H(\omega\subs{m} - \omega\subs{c})
  \right]
\Bigl)
\, \lvert H(\omega\subs{m}) \rvert^2
\label{eq:f-blds}
\end{multline}
with $\omega\subs{c} = 2 \pi f\subs{c}$ the cantilever frequency;
$\omega\subs{m} = 2 \pi f\subs{m}$ the modulation frequency;
$k\subs{c}$ the cantilever spring constant;
$V\subs{ts} = \SI{2}{\volt}$ the tip--sample voltage;
$C_q^{\prime\prime} = C\subs{tip}^{\prime\prime} - 2 (C\subs{tip}^{\prime})^2 \big/ C\subs{tip}$ and 
$\Delta C^{\prime\prime} = 2 (C\subs{tip}^{\prime})^2 \big/ C\subs{tip}$, two derivatives derived from the tip--sample capacitance; and 
\begin{equation}
{H}(\omega) = \frac{1/(j \omega C\subs{tip})}{Z(\omega) + 1/(j \omega C\subs{tip})}
\label{eq:H}
\end{equation}
a complex-valued transfer function function that depends on the tip capacitance and sample impedance $Z(\omega)$. 
The transfer function relates the steady-state tip charge $q(\omega)$ to the applied tip--sample voltage, $q = C\subs{tip} {H}(\omega) V\subs{ts}$.

It is convenient to define a sample response time and a tip--charge response time as 
$\tau\subs{s} =R\subs{s} C\subs{s}$ and 
$ \tau\subs{tip} =R\subs{s} C\subs{tip}$, respectively, 
with $R\subs{s}$ the sample resistance, 
$C\subs{s}$ the sample capacitance, and 
$C\subs{tip}$ the tip capacitance. 
The \eqn{eq:H} transfer function can be written in terms of these charge response times as follows:
\begin{equation}
{H}(\omega) = \frac{\tau\subs{s} \omega - j}{(\tau\subs{s}+\tau\subs{tip})\omega - j}.
\label{eq:transfer-tau}
\end{equation}
According to \eqns{eq:H}{eq:transfer-tau}, BLDS spectra have a roll-off frequency and high frequency plateau determined by the time constants $\tau\subs{s}$ and $\tau\subs{tip}$. 
The high frequency plateau, $\tau\subs{s}/(\tau\subs{s} + \tau\subs{tip}) = C\subs{s}/(C\subs{s}+ C\subs{tip})$, gives information about the ratio of sample capacitance to tip capacitance. 
The roll-off frequency is $1/(\tau\subs{s} + \tau\subs{tip}) = 1/R\subs{s} (C\subs{s} + C\subs{tip})$.
At fixed tip--sample separation, if $C\subs{s} \ll C\subs{tip}$ then the conductivity is simply proportional to the roll-off frequency.
This was the case in Reference~\citenum{Tirmzi2017feb}.
In contrast, here we find $C\subs{s} \gg C\subs{tip}$.
In this limit we can nevertheless obtain $R\subs{s}$ by fitting the entire BLDS spectrum.
The \fig{fig:blds-data} data were fit to \eqns{eq:f-blds}{eq:transfer-tau} with $\tau\subs{s}$, $\tau\subs{tip}$, $C^{\prime\prime}_q$, and $\Delta C^{\prime\prime}$ as fit parameters.
All fit parameters at each light intensity can be found in Tables~S4--S16.
 
\begin{figure}
    \centering
    \includegraphics[width=3.3in]{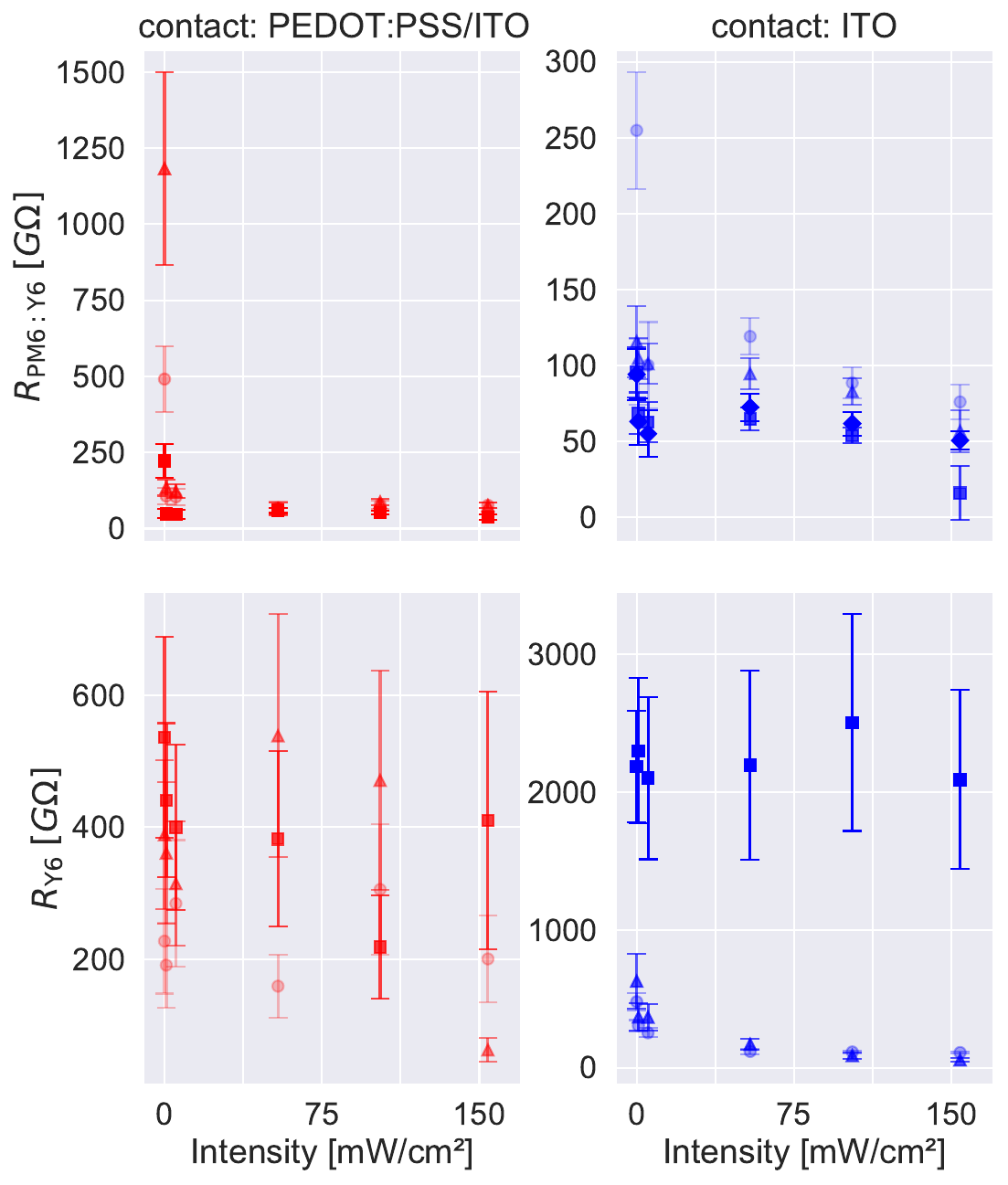}
    \caption{Sample resistance $R_\mathrm{s}$ versus light intensity for all samples. Each symbol corresponds to a different physical sample. Red data points are samples that contain PEDOT:PSS, and blue data points are samples without PEDOT:PSS. Data were collected in triplicate. All error bars are 1$\sigma$.}
    \label{fig:Rs-v-light}
\end{figure}

Sample resistance $R\subs{s}$ was estimated by computing the ratio $\tau\subs{tip} / C\subs{tip}$.
To estimate the tip capacitance, the cantilever tip would usually be modeled as a sphere plus a cone. 
However, Hoepker et al.\ \cite{Hoepker2011dec} showed that the cone capacitance does not contribute significantly to measured friction or frequency noise, so in our analysis we set $C\subs{tip} = C\subs{sphere}$, neglecting $C\subs{cone}$.
We estimate $C\subs{tip} = \SI{4.86}{\atto\farad}$ for a radius $r\subs{tip} = \SI{38.4}{\nano\meter}$ located \SI{120}{\nano\meter} over a semi-infinite ground plane. 
The estimated resistance of the PM6:Y6 and Y6 films is shown as a function of light intensity in \fig{fig:Rs-v-light}.
Values for sample resistance and capacitance are listed in Tables~S17--S20.

\createbibliography

\subsection{Connecting macroscopic resistance to microscopic conductivity}

We wish to compute sample conductivity $\sigma$ from the resistance $R_{\mathrm{s}}$ obtained from BLDS experiments.
Conductivity is an intrinsic quantity that depends on sample and contact materials. 
Conductivity is the product of charge density $\rho$ and charge mobility $\mu$, $\sigma = e \rho \mu$, with $e$ the unit of charge.
Resistance is inversely proportional to sample conductivity,
\begin{equation}
R_{\mathrm{s}} = \frac{1}{k_R \, \sigma},
\label{eq:kR}
\end{equation}
with the proportionality constant, $k_R$, having units of length.
For a slab of area $A$ and length $\ell$, $k_R = A / \ell$.
Because it depends on sample dimensions, the resistance is an extrinsic quantity.

In our experiment, $k_R$ depends on the effective area and thickness of the sample probed by the tip, which in turn depend on tip radius, tip--sample separation, and sample thickness.
Computing $k_R$ for our experiment requires a microscopic theory of the BLDS signal.
Describing the response of free charges and molecular dipoles in the sample to an oscillating tip charge in the BLDS experiment involves coupling a transport equation for free charges with Maxwell equations and the fluctuation--dissipation theorem; this work is beyond the scope of the present study.
Lekkala, Marohn, and Loring, however, have developed a microscopic theory for \emph{friction} over a semiconductor \cite{Lekkala2012sep,Lekkala2013nov}, and Dwyer et al.'s impedance treatment of the tip--sample interaction \cite{Dwyer2019jun}, 
used above to interpret the BLDS spectrum, also yields an expression for cantilever friction.
Let us therefore estimate $k_R$ by comparing the friction predicted by these two treatments.

\begin{table}
\begin{tabular}{r@{\hskip 1em}lcl}
\toprule
 & parameter & symbol &  value \\ \midrule \\
sample & thickness & $h_{\mathrm{s}}$ & \SI{110}{\nano\meter} \\ 
 & dielectric constant & $\epsilon_{\mathrm{s}}$ & 3.4 \\
 & charge density & $\rho$ & \SI{e17}{} to \SI{e27}{\meter^{-3}} \\
 & charge mobility & $\mu$ & \SI{4.0e-4}{\centi\meter^2 \volt^{-1} \second^{-1}} \\ \\
 cantilever & resonance frequency & $\omega$ & $2 \pi \times \SI{75}{\kilo\hertz}$ \\
  & tip radius & $r\subs{tip}$ & \SI{38.4}{\nano\meter} \\
  & tip--sample separation & $h$ & \SI{120}{\nano\meter} \\  
  & tip--sample voltage & $V_{\mathrm{ts}}$ & \SI{1.0}{\volt} \\ \\
\bottomrule
\end{tabular}
\caption{Sample and cantilever parameters used to compute the voltage-normalized cantilever friction using Model I in Reference~\citenum{Lekkala2013nov}.}
\label{eq:Lekkala-sim-input}
\end{table}

\begin{figure}
\centering
\includegraphics[width=2.75in]{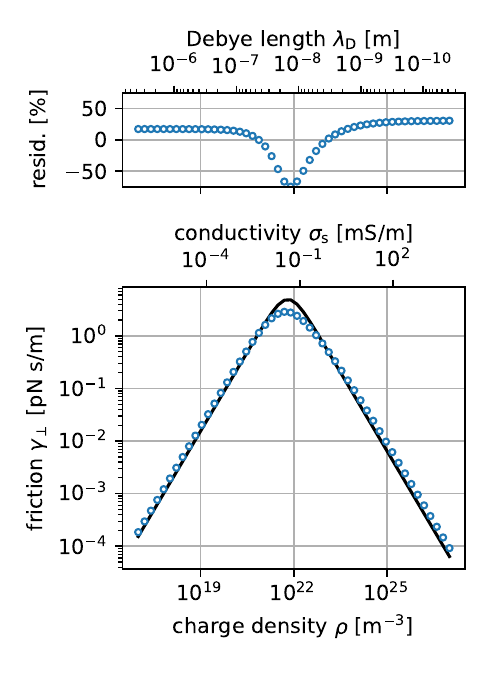}
\vspace{-2em}
\caption{Friction versus charge density.  
Friction was computed numerically using Model I in Reference~\citenum{Lekkala2013nov} and the parameters in Table~\ref{eq:Lekkala-sim-input} (lower, circles).  
The computed friction was fit to \eqn{eq:gamma-Dwyer-empirical} (lower, lines; upper, fit residuals).}
\label{fig:Lekkala-sim-output}
\end{figure}

Friction was calculated for a cantilever oscillating perpendicular to the surface of a semi-infinite semiconductor using Equations (8), (16), and (17) in Reference~\citenum{Lekkala2013nov} and the parameters listed in Table~\ref{eq:Lekkala-sim-input}.
In the language of Reference~\citenum{Lekkala2013nov}, we computed $\gamma_{\perp}$ (noncontact friction) for a Model I sample, with the dielectric constant of the semi-infinite substrate set to $\epsilon_{d} = \SI{e6}{}$ to mimic a metal.
Computations were carried out by \texttt{Numba}-optimized Python code publicly available in the \href{https://github.com/Marohn-Group/dissipationtheory}{\texttt{dissipationtheory}} package; this code was validated by comparing to a low-density analytical expansion and the friction versus\ charge density plots in Figures~7(b) and 9(b) of Reference~\citenum{Lekkala2013nov}.
In \fig{fig:Lekkala-sim-output} we plot the friction versus\ charge density computed for the Table~\ref{eq:Lekkala-sim-input} sample.
The sample mobility was taken to be the average of the electron and hole mobilities given for PM6:Y6 in Table~\ref{tab:mobilities}.
The friction rises at low charge density, reaches a maximum near a charge density of \cite{Lekkala2013nov} $\epsilon_{\mathrm{s}} \epsilon_0 \omega \big/ e \mu = \SI{2.2e21}{\meter^{-3}}$, and decreases at high charge density.
The presence of a friction maximum is in qualitative agreement with Dwyer's model \cite{Tirmzi2017feb,Dwyer2019jun}.

Dwyer et al.\ predict \cite{Dwyer2019jun}
\begin{equation}
\gamma_{\perp}
  = - \frac{C_1^2 V_{\mathrm{ts}}^2 }{\omega \, C_0} 
        \mathrm{Imag}  \! \left[ H(\omega) \right]
  = \frac{C_1^2 V_{\mathrm{ts}}^2 R_{\mathrm{s}}}{1 + \omega^2 R_{\mathrm{s}}^2 (C_0 + C_{\mathrm{s}})^2 }
\label{eq:gamma-Dwyer}
\end{equation}
with $\omega = 2 \pi f_c$ the cantilever frequency, $C_0$ the tip capacitance, $C_1$ the first derivative of the tip capacitance, $V\subs{ts}$ the tip--sample voltage, and $H(\omega)$ the transfer function given by \eqns{eq:H}{eq:transfer-tau}, computed assuming the tip--sample impedance model sketched in \fig{fig:experimental}(c).
Both $C_0$ and $C_1$ were computed from an analytical formula for the capacitance of a sphere over a metallic half plane, Equations~(52) and (53) in Reference~\citenum{Lekkala2012sep} taken in the limit $\epsilon^{\prime}_{\mathrm{rel}}(0) \rightarrow \infty$.

With $R_{\mathrm{s}}$ given by \eqn{eq:kR} and treating $k_R$ and $C_{\mathrm{s}}$ as free parameters, the calculated $\gamma_{\perp}$ versus\ $\rho$ curve in \fig{fig:Lekkala-sim-output} was fit to \eqn{eq:gamma-Dwyer}.
The fit was poor unless $C_{\mathrm{s}}$ was allowed to be negative, which is unphysical.   
In other words, the fit required $C_{\mathrm{total}} = C_0  + C_{\mathrm{s}}$ to be \emph{less} than $C_0$.
Given this observation, let us write the capacitance as 
\begin{equation}
C_{\mathrm{total}} = 4 \pi \epsilon_0 k_C
 \label{eq:kC}
\end{equation}
with the proportionality constant, $k_C$, having units of length.
The \eqn{eq:kC} parameterization is useful because it allows $C_{\mathrm{total}}$ to be directly compared to the capacitance of a sphere of radius $r\subs{tip}$, $4 \pi \epsilon_0 r\subs{tip}$; if $k_C < r\subs{tip}$, then $C_{\mathrm{total}}$ is less than the tip capacitance at infinite tip-sample separation.  
Substituting \eqns{eq:kR}{eq:kC} into \eqn{eq:gamma-Dwyer} we obtain the empirical expression
\begin{equation}
\gamma_{\perp} 
 = \frac{C_1^2 V_{\mathrm{ts}}^2 \sigma k_R}
            {\sigma^2 k_R^2 + \omega^2 (4 \pi \epsilon_0 k_C)^2},
\label{eq:gamma-Dwyer-empirical}
\end{equation}
with $\sigma = e \rho \mu$ the sample conductivity.
The numerically calculated friction in \fig{fig:Lekkala-sim-output} was fit to \eqn{eq:gamma-Dwyer-empirical} assuming proportional errors of one percent.
The fit result is shown as a solid line in \fig{fig:Lekkala-sim-output}(bottom), with the fit residuals displayed above as a percentage error.
The best-fit parameters are 
\begin{subequations}
\begin{align}
k_R^{\mathrm{opt}} & = \SI{44.5}{\nano\meter}, \\
k_C^{\mathrm{opt}} &= \SI{35.1}{\nano\meter}.
\end{align}
\end{subequations}
Considering the simplicity of the \eqn{eq:gamma-Dwyer-empirical} ansatz, the global fit is remarkably good.
Over ten decades of charge density and nearly three decades of friction, \eqn{eq:gamma-Dwyer-empirical} predicts the friction within 17\% at low density and within 30\% at high density.
Conductivity was computed from the BLDS-inferred resistance using
\begin{equation}
\sigma = \frac{1}{k_R^{\mathrm{opt}} R_{\mathrm{s}}}.
\label{eq:sigma}
\end{equation} 

In \fig{fig:Lekkala-sim-output}, the conductivity $\sigma$ is indicated as a second $x$ axis, and the Debye length
\begin{equation}
\lambda_{\mathrm{D}} = \sqrt{\frac{\epsilon_0 k_{\mathrm{b}} T}{e^2 \rho}}
\end{equation}
is shown above the residuals as a third $x$ axis.
We will see below that most experiments were done in the high-conductivity region of \fig{fig:Lekkala-sim-output}.
In this regime, the Debye length is \SI{10}{\nano\meter} or less, consistent with our approximation of treating the sample as a metal when computing $C_0$ and $C_1$.
The short Debye length implies a shallow region of accumulated or depleted charge at the sample surface, which can be modeled as a capacitor operating in parallel with $C_0$, lowering the effective tip capacitance.
The predicted lowering of the tip capacitance is consistent with the observed $k_C^{\mathrm{opt}} < r\subs{tip}$. 

\createbibliography

\subsection{The charge recombination rate can be estimated from conductivity versus light intensity}

 \begin{figure*}
    \includegraphics[width=6.6in]{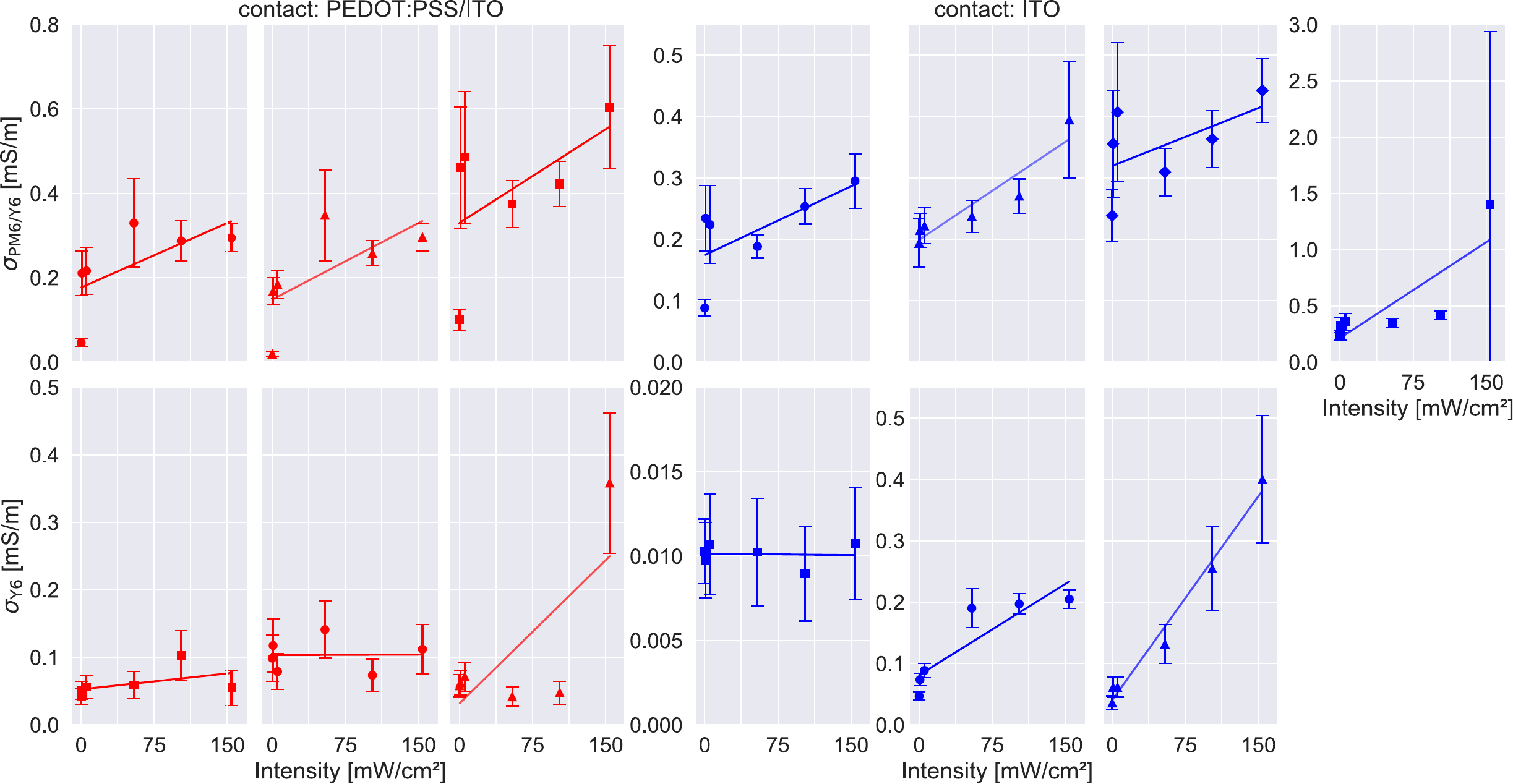}
    \caption{Conductivity $\sigma$ versus\ light intensity for PM6:Y6 samples. 
    Each line/symbol corresponds to a different physical sample. Red data points are samples that contain PEDOT:PSS, and blue data points are samples without PEDOT:PSS. Data were collected in triplicate. All error bars are 1$\sigma$.}
    \label{fig:conductivity}
\end{figure*}    

With conductivity $\sigma$ obtained from \eqn{eq:sigma}, the $\sigma$ versus\ light intensity $I_{h\nu}$ data were plotted and fit to
\begin{equation}
    \sigma = m I_{h\nu} + \sigma_0,
    \label{eq:linear-fit}
\end{equation}
with $m$ a slope and $\sigma_0$ a dark conductivity, \fig{fig:conductivity}.
The dark conductivity $\sigma_0$ is listed in the first column Table~\ref{tab:gamma}. 

\begin{table}
    \begin{center}
    \begin{tabular}{r@{\hspace{3ex}}r@{\hspace{3ex}}r@{\hspace{3ex}}r} \toprule
        sample & $\mu_n [\mathrm{cm^2 V^{-1} s^{-1}}]$ & $\mu_p [\mathrm{cm^2 V^{-1} s^{-1}}]$ & ref.  \\ \midrule
        PM6:Y6 & $1.2 \times 10^{-4}$  & $7.1 \times 10^{-5}$ & \cite{Hosseini2020sep} \\
        PM6:Y6 & $4.8 \times 10^{-4}$ & $5.6 \times 10^{-4}$ & \cite{Yao2021may} \\
        PM6:Y6 & $5.9 \times 10^{-4}$ & $2.0 \times 10^{-4}$ & \cite{Yuan2019apr} \\
        PM6:Y6 & $3.5 \times 10^{-4}$ & $2.5 \times 10^{-4}$ & \cite{Fu2023mar} \\
        PM6:Y6 & $1.2 \times 10^{-3}$ & $2.0 \times 10^{-4}$ & \cite{Li2021dec} \\
        Y6 & $6.5 \times 10^{-4}$ & $1.8 \times 10^{-4}$ & \cite{Yao2021may} \\  \bottomrule
    \end{tabular}
    \end{center}
    \caption{Literature estimates for mobilities used in the calculations below. 
    The electron and hole mobility in PM6:Y6 was taken to be the average of the five reported values.}
    \label{tab:mobilities}
\end{table}

\begin{table*}
    \begin{center}
    \begin{tabular}{c@{\hspace{3ex}}r@{\hspace{2ex}}r@{\hspace{3ex}}r@{\hspace{2ex}}r@{\hspace{3ex}}r@{\hspace{2ex}}r@{\hspace{3ex}}} \toprule
          & \multicolumn{2}{c}{\shortstack{dark conductivity \\ $\sigma_0$ [mS/m]}} 
          & \multicolumn{2}{c}{\shortstack{reduction factor \\ $\gamma$ [unitless]}}  
          & \multicolumn{2}{c}{\shortstack{background charge density \\ $\rho_0 \times 10^{22}$ [m$^{-3}$]}} \\  \cmidrule(rl){2-3} \cmidrule(rl){4-5} \cmidrule(rl){6-7}
      dataset 
         & \footnotesize{PEDOT:PSS/ITO} 
         & ITO 
         & \footnotesize{PEDOT:PSS/ITO}
         & ITO 
         & \footnotesize{PEDOT:PSS/ITO}  
         & ITO \\ \midrule
      $\circ$      & $0.177 \pm 0.048$ & $0.174 \pm 0.032$ & $0.497 \pm 0.230$ & $0.688 \pm 0.308$ &  $1.37 \pm 0.37$ & $1.36 \pm 0.25$ \\ 
      $\triangle$ & $0.148 \pm 0.053$ & $0.199 \pm 0.016$ & $0.500 \pm 0.208$ & $ 0.183 \pm 0.063$ & $1.15 \pm 0.92$ & $1.55 \pm 0.12$ \\
      $\square$  & $0.330 \pm 0.086$ & $0.213\pm 0.150$ & $0.183 \pm 0.110$ & $0.074 \pm 0.040$ &  $2.56 \pm 0.67$  & $1.65 \pm 1.17$ \\
      $\diamondsuit$ & --- & $0.320 \pm 0.037$ & ---  & $0.446 \pm 0.300$ & --- & $2.28 \pm 0.29$  \\ \bottomrule
    \end{tabular}
    \end{center}
    \caption{Measured dark conductivity $\sigma_0$, estimated Langevin reduction factor $\gamma$,  and background charge density $\rho_0$ for each PM6:Y6 sample. 
    Symbols in the dataset column correspond to the symbols in \fig{fig:conductivity}.}
    \label{tab:gamma}
\end{table*}

\begin{figure}
    \centering
    \includegraphics[width=2.25in]{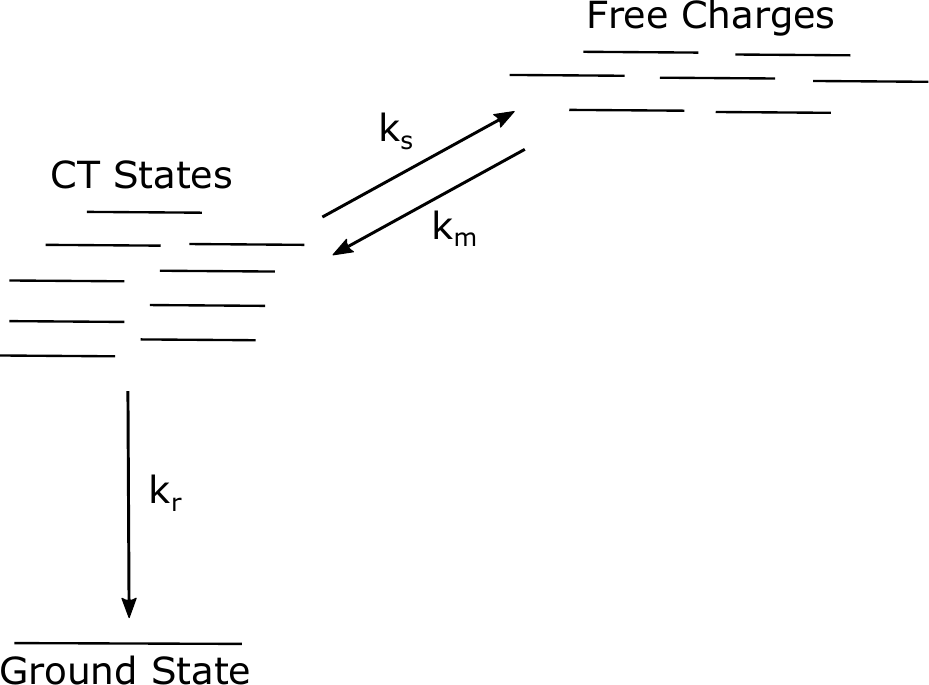}
    \caption{Kinetic scheme showing the rate of charges splitting $k\subs{s}$, meeting $k\subs{m}$, and recombining $k\subs{r}$ in an organic solar cell. 
    Figure adapted from Reference~\citenum{Burke2015apr}.}
    \label{fig:kinetic-scheme}
\end{figure}

From the \fig{fig:conductivity} data we can estimate the charge recombination rate and compare it to Langevin theory. 
We model the carriers in our sample following Burke et al. \cite{Burke2015apr}, who posited an equilibrium between free charges, charge-transfer states, and the ground state.
The associated kinetic scheme is sketched in \fig{fig:kinetic-scheme}.
The coupled equations governing species concentrations, including background or dark carriers, are
\begin{subequations}
\begin{align}
\begin{split}
    \dot{n}&= -k\subs{m}(n_0+\Delta n) \Delta p + G_{h\nu} + k\subs{s}n\subs{CT},
    \label{eq:rate-a}
\end{split}\\
\begin{split}
    \dot{p} &= -k\subs{m}(n_0+\Delta n) \Delta p + G_{h\nu} + k\subs{s}n\subs{CT},
    \label{eq:rate-b}
\end{split}\\
\begin{split}
    \dot{n}\subs{CT} &= k\subs{m}(n_0+\Delta n) \Delta p - k\subs{s}n\subs{CT} - k\subs{r}n\subs{CT},
    \label{eq:rate-c}
\end{split}
\end{align}
\end{subequations}
where $n_0$ and $p_0$ are the background electron and hole concentrations, respectively;
$\Delta n$ and $\Delta p$ are the change in electron and hole concentrations due to light;
$n\subs{{CT}}$ is the charge transfer (CT) state density;
$k\subs{m}$, $k\subs{s}$, and $k\subs{r}$ are the rates at which carriers meet, split, and recombine, respectively;
and $G_{h\nu}$ is the charge-generation rate, proportional to the irradiation intensity and absorption coefficient.
A more detailed explanation of how we arrived at these equations is given in Section~S-4 of the SI. 
At steady state, $\dot{n} = \dot{p} = \dot{n}\subs{CT} = 0$.
Comparing \eqns{eq:rate-a}{eq:rate-c}, we see that
\begin{equation}
    G_{h\nu} = k\subs{r} n\subs{CT}
    \label{eq:G}
\end{equation}
at steady state.  
Plugging \eqn{eq:G} into \eqn{eq:rate-a} and setting the result equal to zero, since we are at steady state, yields
\begin{equation}
    (n_0 + \Delta n) \Delta p = \frac{G_{h\nu}}{\gamma k\subs{{L}} },
    \label{eq:steady-state}
\end{equation}
where $k\subs{{L}} = k\subs{m}$ is the Langevin rate, the rate at which carriers meet, and 
$\gamma = k\subs{r}/(k\subs{r}+k\subs{s})$ is the Langevin-reduction factor in the Burke picture \cite{Burke2015apr}. 

The conductivity is given by 
\begin{equation}
  \sigma = q \mu_n (n_0 + \Delta n) + q \mu_p \Delta p.
  \label{eq:sigma-carriers}
\end{equation}
We see in \fig{fig:conductivity} and Table~\ref{tab:gamma} that $\sigma_0$ is non-zero, from which we conclude that background carriers are present in the dark. 
Let us assume for simplicity the background electron charge density $n_0$ is non-zero and that the background hole density $p_0$ is much less than the change in hole concentration due to light $\Delta p$, i.e.,\ $p_0 \ll \Delta p$. 
In the limit that   $\Delta n \ll n_0$, we can rearrange \eqn{eq:steady-state} to obtain
\begin{equation}
    \Delta p = \frac{1}{n_0}\frac{G_{h\nu}}{\gamma k\subs{{L}}}.
    \label{eq:sigma-langevin-a}
\end{equation}
In this limit
\begin{equation}
  \sigma \approx q \mu_n n_0 + q \mu_p \frac{G_{h\nu}}{n_0 \gamma k\subs{{L}}} ,   
  \label{eq:sigma-langevin-b}
\end{equation}
with $G_{h\nu} = I_{h\nu} \alpha / E $ the charge-generation rate, 
$I_{h\nu}$ [\si{\watt\per\square\meter}] the light intensity,
$\alpha$ [\si{\per\meter}] the absorption coefficient,  
$E$ [\si{\joule}] the energy per photon, 
and $\mu_n$ ($\mu_p$) [\si{\meter\squared\per\volt\per\second}] the electron (hole) mobility.
\eqn{eq:sigma-langevin-b} predicts a conductivity that is linearly proportional to light intensity, consistent with the \fig{fig:conductivity} experiment.

Comparing \eqns{eq:linear-fit}{eq:sigma-langevin-b} we find
\begin{subequations}
\begin{align}
    m & = q \mu_p \frac{\alpha}{E}\frac{1}{n_0 \gamma k\subs{{L}}}, \label{eq:m} \\
    \sigma_0 & = q \mu_n n_0. \label{eq:sigma_0}
\end{align}
\end{subequations}
We obtained $\gamma k\subs{{L}}$ and $n_0$ from the best-fit $m$ and $\sigma_0$, the measured $
\alpha$, known $E$, and literature estimates for $\mu_p$ and $\mu_n$ (Table~\ref{tab:mobilities}).
From Equations \ref{eq:m} and \ref{eq:sigma_0},
\begin{equation}
    \gamma k\subs{L} = \frac{\alpha}{\sigma_0  \, m} \frac{q^2 \mu_n \mu_p}{ E },
    \label{eq:gamma}
\end{equation}
which can be compared to the Langevin rate \cite{Langevin1903jan}  
\begin{equation}
    k_\mathrm{{L}} = \frac{q}{\epsilon\subs{r} \epsilon_0} (\mu_n + \mu_p)
    \label{eq:kL}
\end{equation} 
with $\epsilon\subs{r}$ the relative dielectric constant (3.5, \cite{Hosseini2020sep}) and 
$\epsilon_0$ the vacuum permittivity.
For PM6:Y6 and Y6, we calculate $k_\mathrm{{L}}$ to be 
\SI{4.16e-16}{\meter\cubed\per\second} and
\SI{4.29e-16}{\meter\cubed\per\second}, respectively.
Dividing Equation (\ref{eq:gamma}) by (\ref{eq:kL}) we get the Langevin reduction factor, 
\begin{equation}
  \gamma 
    = \frac{1}{\sigma_0 m} 
       \frac{\mu_n \mu_p}{\mu_n + \mu_p} 
       \frac{\alpha q \epsilon\subs{r} \epsilon_0 }{E}.
  \label{eq:gamma-actual}
\end{equation}

\createbibliography

\subsection{Estimating the dark conductivity}

Let us estimate the charge density expected in Y6 and PM6 near the ITO and PEDOT:PSS interfaces.
The energy levels in a molecular semiconductor like Y6 and PM6 follow a Gaussian distribution.
The hole and electron density can be computed by multiplying a Gaussian density of states by the Fermi--Dirac distribution and integrating over all possible energies:
\begin{subequations}
\begin{align}
n_{\mathrm{h}} &= \frac{\rho_{\mathrm{molec}}}{\sqrt{2 \pi \sigma^2_{\mathrm{v}}}} 
  \int_{-\infty}^{+\infty}  
    \frac{e ^{-(\varepsilon - \varepsilon_{\mathrm{HOMO}})^2 \big/ 2 \sigma^2_{\mathrm{HOMO}}}}
            {e^{\, \beta (\varepsilon - \mu)} + 1}
     d\varepsilon
\\
n_{\mathrm{e}} &= \frac{\rho_{\mathrm{molec}}}{\sqrt{2 \pi \sigma^2_{\mathrm{e}}}}
  \int_{-\infty}^{+\infty}  
    \frac{e ^{-(\varepsilon - \varepsilon_{\mathrm{LUMO}})^2 \big/ 2 \sigma^2_{\mathrm{LUMO}}}}
            {e^{\, \beta (\mu - \varepsilon)} + 1}
     d\varepsilon
\end{align}
\label{eq:carrier-conc}
\end{subequations}
\noindent In \eqn{eq:carrier-conc}, 
$\beta = (k_{\mathrm{b}} T)^{-1}$, with $k_\mathrm{b}$ Boltzmann's constant and $T$ temperature;
$\rho_{\mathrm{molec}}$ is the molecular density, 
($\varepsilon_{\mathrm{LUMO}}$, $\sigma_{\mathrm{LUMO}}$) and ($\varepsilon_{\mathrm{HOMO}}$, $\sigma_{\mathrm{HOMO}}$) are the mean energy and energetic disorder of the lowest unoccupied and highest unoccupied molecular orbitals, respectively; and 
$\mu$ is the electron chemical potential in the molecular film. 
Relevant parameters are listed in Table~\ref{table:levels}.
The table also lists the Fermi level, $\varepsilon_{\mathrm{F}}$.
For simplicity we have assumed $\sigma_{\mathrm{LUMO}} = \sigma_{\mathrm{HOMO}} = \sigma$.
In this limit, $\varepsilon_{\mathrm{F}} = (\varepsilon_{\mathrm{HOMO}} + \varepsilon_{\mathrm{LUMO}})/2$ for undoped Y6 and PM6. 

\begin{table}
\begin{tabular}{r@{\hskip 2em}S@{\hskip 1em}S@{\hskip 1em}S@{\hskip 1em}S}
\toprule
property & {Y6} & {PM6} & {ITO} & {PEDOT:PSS} \\ 
\midrule
$\varepsilon_{\mathrm{LUMO}}$ [eV]  & -4.1 & -3.5  \\
$\varepsilon_{\mathrm{F}}$  [eV]          & -4.88 & -4.53 & -4.70 & -5.02 \\
$\varepsilon_{\mathrm{HOMO}}$ [eV] & -5.65 & -5.56  \\
$\sigma$ [meV]                                       & 63 & 70 \\ 
$\rho_{\mathrm{molec}}$ [$\mathrm{nm}^{-3}$]               & 23.3         & 23.3 \\
\bottomrule
\end{tabular}
\caption{Literature energy-level parameters and molecular density for Y6 and PM6 and Fermi level for ITO and PEDOT:PSS \cite{Kim2019nov,Lv2021may,Tang2018nov,Perdigon-Toro2022mar,Hosseini2020sep,Perdigon-Toro2020jan,Hosseini2023jan}. See Tables~S1 and S2 for individual references. $\rho_{\mathrm{molec}}$ was calculated based on Reference \citenum{Xiang2022dec}.} 
\label{table:levels}
\end{table}

When a Y6 or PM6 molecule is brought near a contact, the molecule and contact will reach a common chemical potential or Fermi level.
To assess the sign of the charge transfer, it is helpful to compute
$\Delta \mu 
    = \varepsilon_{\mathrm{F}}^{\mathrm{contact}} 
    - \varepsilon_{\mathrm{F}}^{\mathrm{molecule}}$.
When $\Delta \mu$ is positive, electrons will flow from the contact to the molecule, whereas
when $\Delta \mu$ is negative, electrons flow from the molecule to the contact.
Charge densities $n_{\mathrm{h}}$ and $n_{\mathrm{e}}$ were computed from Equations~\ref{eq:carrier-conc} with $\mu \rightarrow \varepsilon_{\mathrm{F}}^{\mathrm{contact}}$, using the parameters in Table~\ref{table:levels}, and a total 
charge density $\rho = n_{\mathrm{h}} - n_{\mathrm{e}}$ computed.
Results are summarized in Table~\ref{table:calc-charge-density}.

\begin{table}
\begin{tabular}{llrr}
\toprule
molecule &      contact & $\Delta \mu$ [mV] & $\rho$ [m$^{-3}$] \\
\midrule
     Y6 &        ITO &            $175$ &   \SI{-3.4e+19}{} \\
    PM6 &        ITO &           $-170$ &    \SI{3.2e+15}{} \\
     Y6 &  PEDOT:PSS &           $-145$ &    \SI{1.1e+19}{} \\
    PM6 &  PEDOT:PSS &           $-490$ &    \SI{7.7e+20}{} \\
\bottomrule
\end{tabular}
\caption{Computed chemical potential difference $\Delta \mu$ and charge density $\rho$ for various organic--metal contacts.}
\label{table:calc-charge-density}
\end{table}

\section{Discussion}

PM6:Y6 was found to be highly conductive, even in the dark.
The conductivity increased linearly with light intensity, on both PEDOT:PSS/ITO and ITO, indicative of PM6:Y6 being a good solar cell material. 

All Y6 samples had much lower conductivity than the blend, indicating that the blend is a better solar cell material, as expected. 
Some Y6 samples showed modest photoconductivity, while in most samples the conductivity was dominated by background carriers.
The conductivity was highly variable, with one sample showing almost no conductivity; this lack of conductivity might be due to material degradation, since the reagents were older as that sample was prepared about two months after the other samples. 
These observations are in qualitative agreement with Sa\u{g}lamkaya et al., who showed that charge generation readily occurs in neat Y6 \cite{Saglamkaya2023mar}. 

The PM6 BLDS spectra were independent of modulation frequency.
The conductivity roll-off must therefore be outside our measurement limits, either below \SI{100}{\hertz} or above \SI{3}{\mega\hertz}.
Based on the moderate conductivity observed in the PM6:Y6 blend, it seems unphysical that the PM6 control would have higher conductivity than the blend.
We conclude that $\sigma\subs{PM6}$ must be less than the lowest conductivity measured in our experiment, \SI{0.0098}{\milli\siemens\per\meter}. 

Values obtained for $\gamma$ in each PM6:Y6 sample are listed in Table~\ref{tab:gamma}. 
For PM6:Y6/PEDOT:PSS/ITO, we find $\gamma = 0.18$ to $0.50$ and $\gamma k\subs{L} = 0.76$ to \SI{2.1e-16}{\meter\cubed\per\second}.
These values are in order-of-magnitude agreement with 
$\gamma k\subs{L} = \SI{2.9e-17}{\meter\cubed\per\second}$ obtained by 
Hosseini et al.\ \cite{Hosseini2020sep} for a Ag/PDINO/PM6:Y6/PEDOT:PSS/ITO solar cell using bias-assisted charge extraction measurements and 
$\gamma k\subs{L} = \SI{2e-17}{\meter\cubed\per\second}$ obtained by Zhang et al.\ \cite{Zhang2021jan} in a Ag/PFNDI/PM6:Y6/PEDOT:PSS/ITO solar cell using transient photovoltage and photocurrent measurements.

\begin{figure}
    \centering
    \includegraphics[width=2.25in]{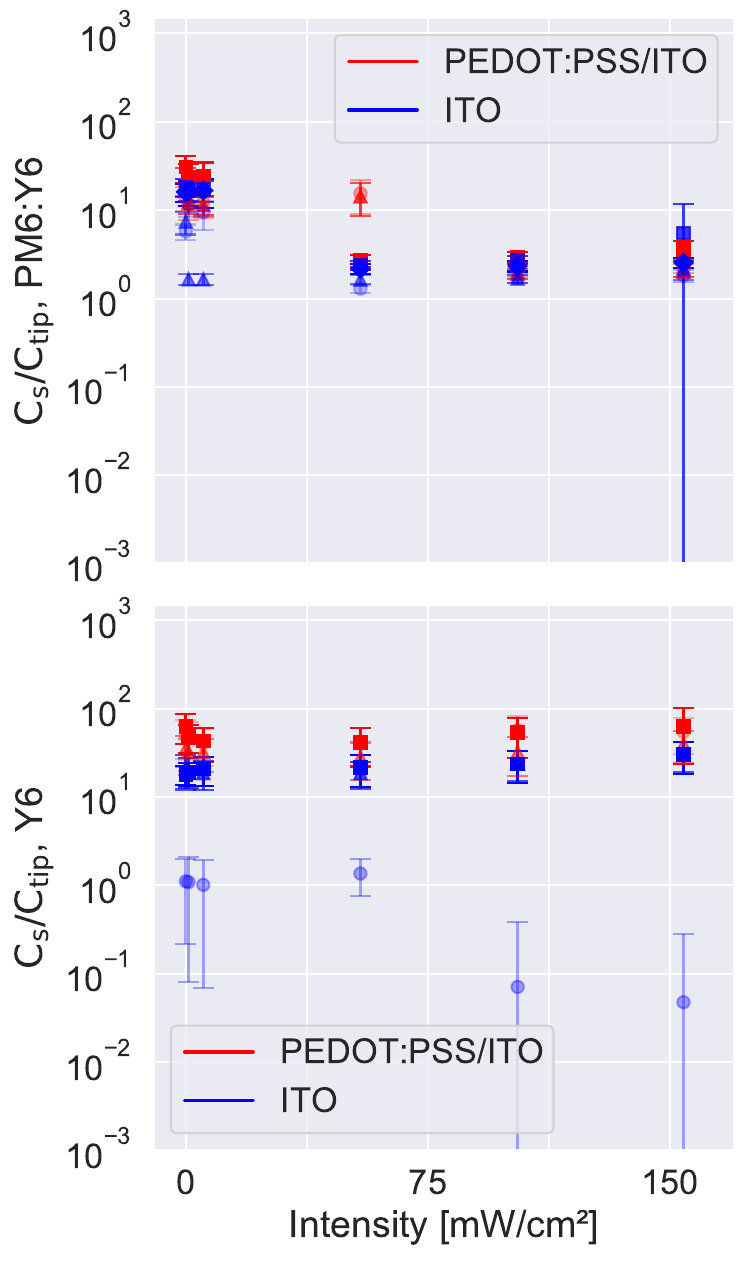}
    \caption{ $C\subs{s} \big/ C\subs{tip}$ for PM6:Y6 and Y6 with PEDOT:PSS/ITO and ITO only contacts. Each symbol corresponds to a different physical sample. Red data points are samples that contain PEDOT:PSS, and blue data points are samples without PEDOT:PSS. Data were collected in triplicate. All error bars are 1$\sigma$.}
    \label{fig:cs-over-ctip}
\end{figure}

The large random error in our determination of $\gamma$ arises primarily from the relatively poor fits of the BLDS spectra.
In our fits we assumed that the sample dielectric constant was frequency-independent, which is likely a poor approximation for the small molecule and polymeric semiconductors studied here.
To determine the proportionality constant connecting the best-fit bulk resistance to a microscopic conductivity, we compared a macroscopic theory to a microscopic theory for friction, and assumed that the same proportionality constant held in a BLDS measurement.
Future work should focus on developing a proper microscopic theory for the BLDS measurement that incorporates a frequency-dependent dielectric constant.

We observed that $C\subs{s} \big/ C\subs{tip}$ was independent of light intensity, as seen in \fig{fig:cs-over-ctip}.
Given that all experiments were carried out at fixed tip--sample separation, $C\subs{tip}$ was constant.
We conclude that $C\subs{s}$ was must also be independent of light intensity.
Since $\tau\subs{s}$ decreased with increasing light intensity, $R\subs{s}$ must be likewise decreasing.
In the organic photovoltaic films studied here, light changed $R\subs{s}$, not $C\subs{s}$ as is universally assumed in EFM experiments \cite{Coffey2006aug,Dwyer2017jun,Karatay2016may}. 
The standard description of EFM simply ignores the sample resistance.
By acquiring BLDS spectra and interpreting them using an impedance model of the tip--sample interaction, we were able to quantify both the resistance and capacitance of our thin-film sample.

We found that if data were collected more than a week after PM6:Y6 or Y6 samples were prepared and stored in a nitrogen box in the dark, they did not display the same high conductivity as was measured immediately after preparation; PM6:Y6 and Y6 samples need to be measured within a day or two after being prepared. 
Consistent with this observation, Zhu et al. found that nonfullerne acceptors can aggregate in bulkheterojunction films, decreasing device performance, just by storing in the dark in a nitrogen-filled glove box \cite{Zhu2019apr}.

Although every precaution was taken to ensure that sample preparation was consistent, we saw distinct differences in the dependence of conductivity on light intensity, \fig{fig:conductivity}.
The reproducibility of organic photovoltaic devices between labs, and within labs, remains poor \cite{TremoletdeVillers2009nov,Chien2017mar,Goetz2022may}.
Our study suggests that BLDS measurements can be a useful non-destructive tool for monitoring, and perhaps improving, the consistency of film conductivity prior to fabricating a full solar cell.

\createbibliography

\section{Conclusions}

We report new evidence for significant dark conductivity in PM6:Y6 and corroborate recent studies indicating non-Langevin recombination in this material.
The underlying concentration of dark carriers is many orders of magnitude larger than expected from Fermi-level equilibration.
Sources of the anomalous dark carriers could be chemically oxidized/reduced PM6 or Y6 present due to reactions of these molecules with air or impurities present due to imperfect synthesis, or interface dipoles absent from our model of Fermi-level equilibration. 

That the observed dark conductivity is comparable to the light-induced conductivity implies that either the electron or hole pseudo-Fermi level is partially pinned by the associated dark carriers in PM6:Y6.
Such pseudo-Fermi level pinning due to dark carriers is a potential source of voltage loss and deserves further study.

\createbibliography

\section{Methods}

\sloppy

The polymer donor PM6 (Ossila) 
and the molecular acceptor Y6 (Ossila) were stored in a nitrogen glove box and used as received within 8 months of receipt.
A small amount was removed from the glove box to prepare each solution in air. 
Chloroform (Macron Fine Chemicals) and 1-chloronaphthalene (Sigma-Alrdrich) were used as received. 

PM6:Y6 samples were prepared on both ITO and PEDOT:PSS/ITO substrates as follows. 
The PEDOT:PSS preparation was adapted from Reference~\citenum{Raab2023may}, with spin coating done statically instead of dynamically. 
The PM6:Y6 preparation followed Reference~\citenum{Perdigon-Toro2020jan} as closely as possible. 
ITO coated glass slides (10 $\Omega$/sq., Nanocs) were cleaned by sonicating in isopropanol and acetone (1:1 volume ratio) for 10 minutes, followed by scrubbing with detergent (Aquet Liquid Laboratory Detergent) in DI water, rinsing with DI water, and drying with $\mathrm{N}_2$. 
Slides were then UV-ozone cleaned (UVO-Cleaner Model No. 12, Jelight Company Inc.) for 10 minutes.  
PEDOT:PSS (Al 4083, Ossila) was filtered through a \SI{0.45}{\micro\meter} PTFE filter and 
\SI{100}{\micro\liter} was statically spin coated at 4000~rpm for \SI{30}{\second}.
Films were annealed on a hot plate at \SI{150}{\celsius} for \SI{20}{\minute} in air and left in a nitrogen flow box to cool before spin coating the active layer.
PM6:Y6 (1:1.2 weight ratio, \SI{16}{\milli\gram\per\milli\liter}) in chloroform:1-chloronaphthalene (CF:CN, 99.5:0.5 volume ratio) was stirred for \SI{3}{\hour} using a new PTFE stir bar prior to statically spin coating \SI{100}{\micro\liter} at 3000~rpm for \SI{60}{\second}.
Control samples were prepared with PM6, \SI{7.3}{\milli\gram\per\milli\liter} in CF:CN, and 
Y6, \SI{8.7}{\milli\gram\per\milli\liter} in CF:CN.  
All solutions were stirred in air in the dark.
Three to four replicate samples were made for each formulation. 
Samples were kept in the dark in a nitrogen box until they were loaded into the microscope under red light. Sample thickness, measured by profilometry (Tencor AlphaStep 500), was \SI{110}{\nano\meter}, \SI{40}{\nano\meter}, and \SI{80}{\nano\meter} for PM6:Y6, Y6, and PM6, respectively.

Scanning probe measurements were performed under high vacuum (\SI{e-5}{\milli\bar}) in a custom-built scanning Kelvin probe microscope. 
The cantilever used (HQ:NSC18/Pt conductive probe, MikroMasch) had a typical 
resonance frequency $f\subs{c} = \SI{75}{\kilo\hertz}$,
force constant $k \subs{c}= \SI{2.8}{\newton\per\meter}$, 
tip radius $r\subs{tip} = \SI{30}{\nano\meter}$, and 
cone angle $\theta\subs{cone} = \SI{40}{\degree}$.
Specific cantilever parameters used in each \fig{fig:blds-data} experiment are given in Table~S3.
Data were collected at a tip--sample separation of \SI{120}{\nano\meter}. 
Cantilever motion was detected using a fiber optic interferometer operating at $\lambda = \SI{1313}{\nano\meter}$ (Corning model SMF-28 Ultra fiber; Applied Optoelectronics Inc.\ model DFB-1310-BF-10-A3-FA laser; New Focus Model 2053-FC photodetector).  
The sample was illuminated from above with a variable intensity continuous wave $\lambda = \SI{639.7}{\nano\meter}$ diode laser (QPhotonics QFLD-635-30SAX), and the estimated intensity at the sample was \SI{0}{} to \SI{154}{\milli\watt\per\square\centi\meter} (details on the laser spot size measurement are given in Section~S-2 in the SI).

The cantilever was driven into self oscillation via positive feedback to an amplitude of \SI{100}{} to \SI{200}{\nano\meter}. 
Tip--sample separation was determined by first approaching the sample surface until the amplitude decreased to 80\% of its initial value and then backing up to the desired separation, \SI{120}{\nano\meter}, using a Thorlabs piezo controller.
Tip--sample separation was checked before and after each BLDS spectrum was collected. 
If the tip--sample separation drifted more than \SI{3}{\nano\meter}, the minimum DC step in our microscope, the spectrum was discarded. 

In a BLDS measurement, the tip voltage was sinusoidally on--off modulated at a fixed frequency, \SI{20}{\hertz}, and sinusoidally modulated at frequencies $\omega_\mathrm{m} = 2\pi f_\mathrm{m}$ ranging from $f_\mathrm{m} = \SI{100}{\hertz}$ to \SI{3}{\mega\hertz} at various light intensities.
The cantilever frequency was measured using a commercial phase-locked loop (RHK Technology PLLPro), and the \SI{20}{\hertz} Fourier component of the cantilever frequency, $\Delta f_\mathrm{BLDS}$, was obtained from the phase-locked loop output using lock-in detection (Perkin Elmer Instruments 7265 DSP Lock-in Amplifier). 
The lock-in output $\Delta f_\mathrm{BLDS}$ was divided by $V_\mathrm{ts}^2$, with $V_\mathrm{ts} = 2$ V, and the resulting voltage-normalized frequency shift was plotted versus\ the modulation frequency $\omega_\mathrm{m}$ to give a BLDS spectrum. 

\createbibliography

\begin{acknowledgements}
Research primarily supported by the National Science Foundation under Award DMR-2113994 (studies of organic semiconductors, electric force microscope development) 
with additional funding from the U.S.\ Department of Energy, Office of Science, Basic Energy Sciences, under Award DE-SC0022305 (electric force microscope development and calibration).
The authors acknowledge the use of facilities and instrumentation supported by the National Science Foundation through the Cornell University Materials Research Science and Engineering Center DMR-1719875.
Software used in this work includes, in part, h5py \cite{Collette2023jan}, lmfit \cite{Newville2023jul}, Matplotlib \cite{Hunter2007jun}, Numba \cite{Lam2015nov}, NumPy \cite{vanderWalt2011mar}, pandas \cite{McKinney2010jun, pandas2024jan}, Pint, SciPy \cite{Virtanen2020mar}, and SymPy \cite{Meurer2017jan}.
\end{acknowledgements}

\section*{Author Contributions}

R.L.C. carried out the experiments and data analysis. 
C.A.P. designed the microscope probe head, implemented microscope improvements, and measured the visible light spot size. 
V.E.M. implemented microscope improvements. 
J.A.M. executed friction simulations and conceived and directed the project. 
R.L.C., C.A.P., and J.A.M. wrote the manuscript, and all authors reviewed the manuscript.

\bibliography{main}

\clearpage

\onecolumngrid
\section*{Table of Contents}

\begin{figure}[h!]
    \centering
    \includegraphics[width=\textwidth]{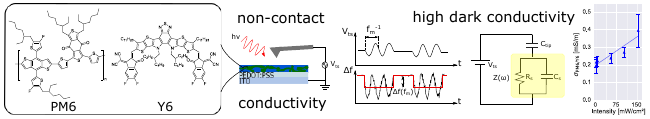}
    \label{fig:ToC}
\end{figure}

Sub-Langevin recombination rates were measured in PM6:Y6 using a non-contact measurement of conductivity in the dark and under illumination. 
An impedance model and a microscopic model of cantilever friction were compared to obtain a microscopic sample conductivity.
Dark conductivity was orders of magnitude higher than expected, suggesting that dark carriers may be a source of open-circuit voltage loss.

\end{document}


\sloppy

\title{Supporting Information:\\Electrical Scanning Probe Microscope Measurements Reveal\\ 
Surprisingly High Dark Conductivity in Y6 and PM6:Y6\\ and 
Non-Langevin Recombination in PM6:Y6}
\author{Rachael L. Cohn}
\affiliation{Department of Chemistry and Chemical Biology, Cornell University, Ithaca, New York 14853 USA}
\author{Christopher A. Petroff}
\affiliation{Department of Chemistry and Chemical Biology, Cornell University, Ithaca, New York 14853 USA}
\affiliation{Department of Materials Science and Engineering, Cornell University, Ithaca, New York 14853 USA}
\author{Virginia E. McGhee}
\affiliation{Department of Chemistry and Chemical Biology, Cornell University, Ithaca, New York 14853 USA}
\author{John A. Marohn}
\affiliation{Department of Chemistry and Chemical Biology, Cornell University, Ithaca, New York 14853 USA}
\date{February 23, 2024}

\maketitle

\clearpage

\tableofcontents

\clearpage

\listoffigures

\clearpage

\listoftables

\clearpage

\section{Full Chemical Names}
\vspace{0.1in}
\noindent PM6:\\ Poly[(2,6-(4,8-bis(5-(2-ethylhexyl-3-fluoro)thiophen-2-yl)-benzo[1,2-b:4,5-b’]dithiophene))-alt-(5,5-(1’,3’-di-2-thienyl-5’,7’-bis(2-ethylhexyl)benzo[1’,2’-c:4’,5’-c’]dithiophene-4,8-dione)]

\vspace{0.1in}
\noindent Y6:\\ 2,2'-((2Z,2'Z)-((12,13-Bis(2-ethylhexyl)-3,9-diundecyl-12,13-dihydro-[1,2,5]thiadiazolo[3,4-e]thieno-[2'',3'':4',5']thieno[2',3':4,5]pyrrolo[3,2-g]thieno-[2',3':4,5]thieno[3,2-b]indole-2,10-diyl)bis(methanylylidene))-bis(5,6-difluoro-3-oxo-2,3-dihydro-1H-indene-2,1-diylidene))dimalononitrile

\vspace{0.1in}
\noindent PEDOT:PSS:\\ Poly(3,4-ethylenedioxythiophene) polystyrene sulfonate

\clearpage
\section{Visible Laser Spot Size and Light Intensity}\label{sec:laser}
\begin{figure}[h]\centering
\includegraphics[width=0.65\columnwidth]{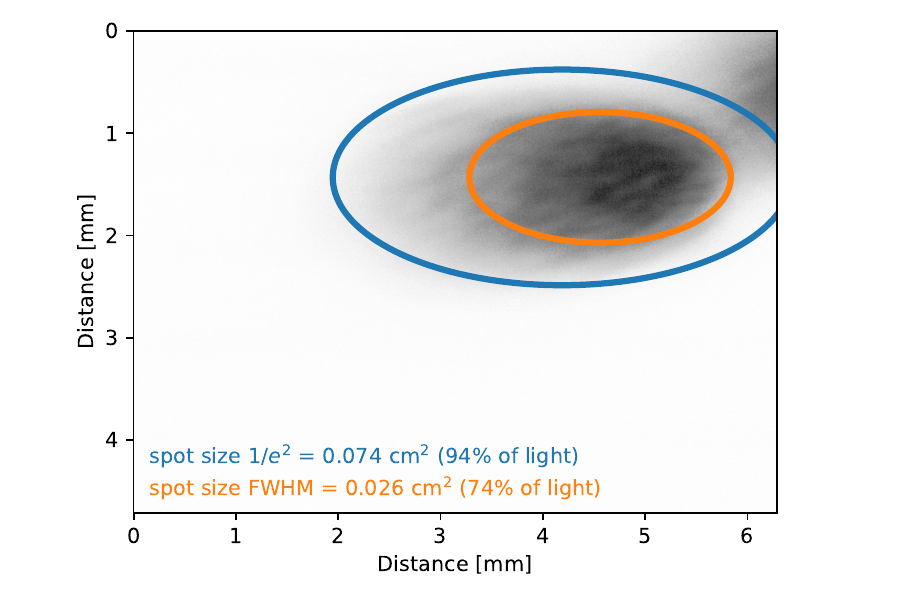}
\vspace{-0.25in}
\caption[Image of the laser spot used to illuminate the sample.]{Image of the laser spot used to illuminate the sample showing the $1/e^2$ (shown in \colorindicator{tab:blue}{blue}) and full width at half max (FWHM) (shown in \colorindicator{tab:orange}{orange}) spot sizes. The FWHM spot size was used to determine the incident light intensity.}
\label{fig:laser-spot}
\end{figure}

The samples were illuminated from above using a variable intensity fiber-coupled red $\lambda = \SI{639.7}{\nano\meter}$ continuous wave diode laser (QPhotonics QFLD-635-30SAX).
The laser output was directed to the sample through a \SI{50}{\micro\meter} core diameter, \SI{0.22}{NA} multimode fiber (Thorlabs FG050LGA).
The cleaved end of the fiber was placed \SI{11}{\milli\meter} away from the sample and directed toward the cantilever tip at a \SI{23}{\degree} angle relative to the sample surface.
The laser spot size was measured using a CMOS sensor, utilizing a method similar to the literature \cite{BonnettDelAlamo2021jul}.
In brief, a Raspberry Pi camera module (Seeed Studio model 114992442 with a Sony IMX477 CMOS sensor) was loaded and approached as a sample would be.
With the laser output set to \SI{13}{\micro\watt}, an image of the laser spot (Fig.~\ref{fig:laser-spot}) was captured and saved as an array.
The \mbox{lmfit} Python package \cite{Newville2023jul} was used to fit the short axis of the elliptical spot to a Gaussian and the long axis (due to the \SI{23}{\degree} angle of incident light) to a skewed Gaussian.
The SciPy Python package's \cite{Virtanen2020mar} signal module was used to determine both the $1/e^2$ width and the full width at half max (FWHM) of each fit.
The percent of the total light contained within each spot was estimated by integrating the fits using the Simpson function from SciPy's integrate module.
We chose to use the FWHM spot size (\SI{0.026}{\centi\meter\squared}) for our light intensity calculations as we expect that the cantilever falls near the center of the laser beam.
The laser light intensity was estimated by multiplying the measured laser power (Coherent FieldMate 1098297 Laser Power Meter with OP-2 VIS 1098313 Si sensor) by the relative area under the curve for FWHM and dividing by the FWHM ellipse spot area.

\clearpage
\section{Absorption Coefficient}

\begin{figure}[h!]
    \begin{adjustwidth}{-1in}{-1in}
    \centering
    \includegraphics[width=6.6in]{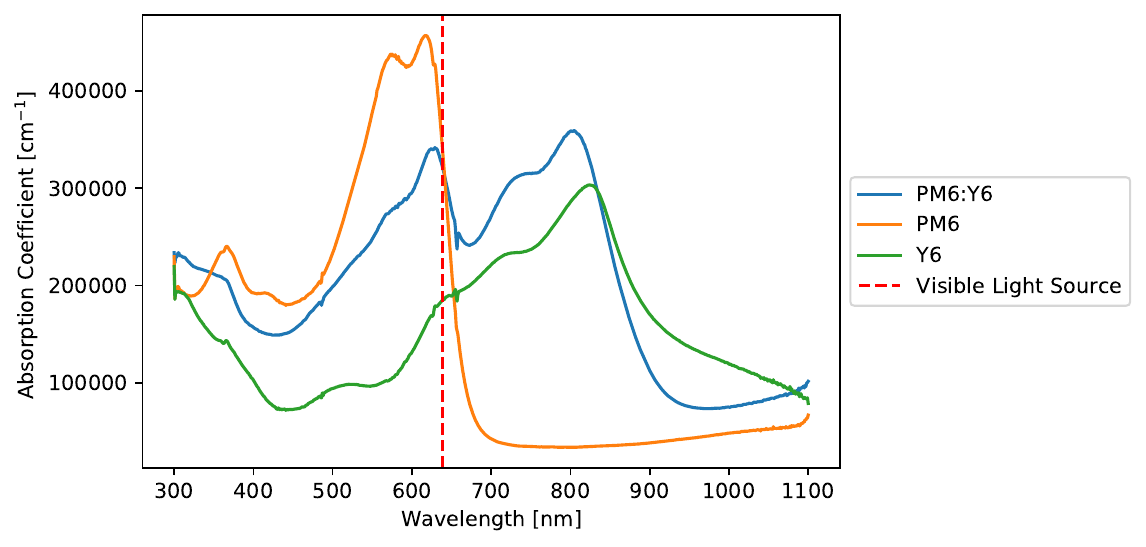}
    \end{adjustwidth}
    \caption[Absorption coefficient spectra.]{Absorption coefficient spectra for PM6:Y6, Y6, and PM6 calculated from the measured UV-Vis spectra.}
    \label{fig:uv-vis}
\end{figure}

UV-Vis spectra of polymer films (Fig.~\ref{fig:uv-vis}) were collected using an Agilent Technologies Cary 8454 UV-Vis. Absorption coefficients match those found in literature within 25\% \cite{Nyman2021mar, Hosseini2020sep}. 

\clearpage
\section{Charge Density Dependence on Generation Rate} \label{sec:charge-density}

\begin{table*}[h!]
    \begin{center}
    \begin{tabular}{c@{\hspace{3ex}}c@{\hspace{3ex}}c@{\hspace{3ex}}c@{\hspace{3ex}}c} \toprule
        Chemical & $\varepsilon_\mathrm{HOMO}$ [eV] & $\varepsilon_\mathrm{LUMO}$ [eV] & Fermi Level [eV] & Ref. \\ 
        \midrule    
        ITO & --- & --- & -4.70 & \cite{Kim2019nov} \\
        PM6 & -5.56 & -3.5 & -4.53 &\cite{Lv2021may} \\
        Y6 & -5.65 & -4.1  & -4.88 &\cite{Lv2021may} \\
        PEDOT:PSS & --- & --- & -5.02 & \cite{Tang2018nov} \\ \bottomrule
    \end{tabular}
    \end{center}
    \caption[Energy levels]{Energy levels with references found in main text Table IV.}
    \label{tab:energies}
\end{table*}

\begin{table}[h!]
\begin{center}
\begin{tabular}{c@{\hspace{3ex}}c@{\hspace{3ex}}c@{\hspace{3ex}}c} \toprule
     PM6 $\sigma_\mathrm{HOMO}$ [meV] & Y6 $\sigma_\mathrm{LUMO}$ [meV] & Ref. \\
     \midrule
      74 & 60 &\cite{Perdigon-Toro2022mar} \\
      63 & 59 &\cite{Hosseini2020sep} \\
      83 & 71 &\cite{Perdigon-Toro2020jan} \\
      60 & 58 &\cite{Hosseini2023jan} \\
     \bottomrule
\end{tabular}
\end{center}
\caption[Disorder parameter values]{Disorder parameter values used to calculate average disorder found in main text Table IV.}
\label{tab:disorder}
\end{table}

How would charge density at steady state depend on the generation rate, $G$, and therefore light intensity or laser power?

\subsection{$n_0 = 0$, no background charge density}

Following Burke \textit{et al.} \cite{Burke2015apr}, we have the rate equations
\begin{subequations}
\begin{align}
\begin{split}
    \dot{n} &= -k_\mathrm{m}np + G + k_\mathrm{s}n_\mathrm{CT}, 
    \label{eq:ndot}
\end{split}\\
\begin{split}
    \dot{p} &= -k_\mathrm{m}np + G + k_\mathrm{s}n_\mathrm{CT}, 
    \label{eq:pdot}
\end{split}\\
\begin{split}
    \dot{n}_\mathrm{CT} &= k_\mathrm{m}np - k_\mathrm{s}n_\mathrm{CT} - k_\mathrm{r}n_\mathrm{CT}.
    \label{eq:n-ct}
\end{split}
\end{align}
\end{subequations}

\noindent with $n = n_0 + \Delta n$ and $p = p_0 + \Delta p$, where $n_0$ and $p_0$ are the background electron and hole concentration and $\Delta n$ and $\Delta p$ are the change in electron and hole density due to light; $k_\mathrm{m}$, $k_\mathrm{s}$, and $k_\mathrm{r}$ are the rates at which carrier meet, split, and recombine respectively; and the generation rate $G = G_{h \nu} + G_0$, where $G_{h \nu}$ is the generation rate due to light and $G_0 = k_\mathrm{m} n_0 p_0$ is the thermal generation rate. 

Expanding Eqs.~\ref{eq:ndot}--\ref{eq:n-ct}, we have

\begin{subequations}
\begin{align}
\begin{split}
    \dot{n} &= -k_\mathrm{m}(n_0 +\Delta n)(p_0 + \Delta p) + G_{h \nu} + G_0 + k_\mathrm{s}n_\mathrm{CT}, 
    \label{eq:ndot2}
\end{split}\\
\begin{split}
    \dot{p} &= -k_\mathrm{m}(n_0 +\Delta n)(p_0 + \Delta p) + G_{h \nu} + G_0 + k_\mathrm{s}n_\mathrm{CT}, 
    \label{eq:pdot2}
\end{split}\\
\begin{split}
    \dot{n}_\mathrm{CT} &= k_\mathrm{m}(n_0 +\Delta n)(p_0 + \Delta p) - k_\mathrm{s}n_\mathrm{CT} - k_\mathrm{r}n_\mathrm{CT}.
    \label{eq:n-ct2}
\end{split}
\end{align}
\end{subequations}

Cancelling like terms and assuming $n_0 = 0$ and $p_0 << \Delta p$, and Eqs.~\ref{eq:ndot2}--\ref{eq:n-ct2} reduce to

\begin{subequations}
\begin{align}
\begin{split}
    \dot{n} &= -k_\mathrm{m}\Delta n \Delta p + G_{h \nu} + k_\mathrm{s}n_\mathrm{CT}, 
    \label{eq:ndot3}
\end{split}\\
\begin{split}
    \dot{p} &= -k_\mathrm{m}\Delta n \Delta p + G_{h \nu} + k_\mathrm{s}n_\mathrm{CT}, 
    \label{eq:pdot3}
\end{split}\\
\begin{split}
    \dot{n}_\mathrm{CT} &= k_\mathrm{m}\Delta n \Delta p  - k_\mathrm{s}n_\mathrm{CT} - k_\mathrm{r}n_\mathrm{CT}.
    \label{eq:n-ct3}
\end{split}
\end{align}
\end{subequations}

At steady state, the above rate equations are set equal to zero and equate Eq.~\ref{eq:ndot3} and Eq.~\ref{eq:n-ct3},

\begin{equation}
\begin{split}
    G_{h \nu} + k_\mathrm{s} n_\mathrm{{CT}} - k_\mathrm{m} \Delta n \Delta p = \\
    k_\mathrm{s} n_\mathrm{{CT}}  +  k_\mathrm{r} n_\mathrm{{CT}} - k_\mathrm{m}\Delta n \Delta p
\end{split} 
\label{eq:steady-state}
\end{equation}

Solving Eq.~\ref{eq:steady-state} for $G_{h \nu}$, we get

\begin{equation}
    G_{h \nu} = k_\mathrm{r} n_\mathrm{{CT}} .
    \label{eq:G}
\end{equation}

The generation rate $G_{h \nu}$ can be calculated using

\begin{equation}
    G_{h \nu} = \frac{I_{h\nu} \alpha}{E}
\end{equation}

\noindent where $I_{h\nu}$ is the visible light intensity, $\alpha$ is the absorption coefficient, and $E$ is the photon energy. We get $E$ from the wavelength of the visible laser, 639.7 nm, $E = hc / \lambda$ = 1.94 eV. $I_{h\nu}$ is estimated from the laser power and spot size.

We plug Eq.~\ref{eq:G} into Eq.~\ref{eq:steady-state} to get

\begin{equation}
    k_\mathrm{m} \Delta n \Delta p = G_{h \nu} + G_{h \nu} \frac{k_\mathrm{s}}{k_\mathrm{r}}
\end{equation}

Dividing both sides by $k_\mathrm{m}$, we have

\begin{equation}
    \Delta n \Delta p = \frac{G_{h \nu}}{k_\mathrm{m}}\left( 1 + \frac{k_\mathrm{s}}{k_\mathrm{r}} \right) = \frac{G_{h \nu}}{k_\mathrm{m}} \left( \frac{k_\mathrm{r} + k_\mathrm{s}}{k_\mathrm{r}} \right)
\end{equation}

and $\frac{k_\mathrm{r}}{k_\mathrm{r} + k_\mathrm{s}} = \gamma$, where $\gamma$ is the Langevin reduction factor. When $k_\mathrm{s} = 0$ we are in the Langevin limit. Letting $x=\Delta n, \Delta p$ we get

\begin{equation}
    x = \left( \frac{G_{h \nu}}{\gamma k_\mathrm{m}} \right)^{1/2} = \left( \frac{G_{h \nu}}{\gamma k_\mathrm{L}} \right)^{1/2}
    \label{eq:x}
\end{equation}

Note that $k_\mathrm{m}$, the rate at which carriers meet, is equal to $k_\mathrm{{L}}$, the Langevin rate. We know that conductivity $\sigma$ is

\begin{equation}
    \sigma = q \mu_n n + q \mu_p p,
    \label{eq:sigma}
\end{equation}

\noindent with $\mu_n$ and $\mu_p$ the electron and hole mobilities, respectively. 

Plugging Eq.~\ref{eq:x} into Eq.~\ref{eq:sigma}, we find

\begin{equation}
    \sigma = q(\mu_n + \mu_p) \left( \frac{G_{h \nu}}{ \gamma k_\mathrm{L}} \right)^{1/2}.
\end{equation}

\subsection{$n_0 \neq 0$, finite background charge density}

Starting with Eqs.~\ref{eq:ndot2}--\ref{eq:n-ct2}, and assuming $p_0 << \Delta p$, we have

\begin{subequations}
\begin{align}
\begin{split}
    \dot{n} &= -k_\mathrm{m}(n_0 \Delta p + \Delta n \Delta p) + G_{h \nu} + k_\mathrm{s} n_\mathrm{CT}, 
    \label{eq:ndot4}
\end{split}\\
\begin{split}
    \dot{p} &= -k_\mathrm{m}(n_0 \Delta p + \Delta n \Delta p) + G_{h \nu} + k_\mathrm{s} n_\mathrm{CT}, 
    \label{eq:pdot4}
\end{split}\\
\begin{split}
    \dot{n}_\mathrm{CT} &= k_\mathrm{m}(n_0 \Delta p + \Delta n \Delta p) - k_\mathrm{s} n_\mathrm{CT} - k_\mathrm{r} n_\mathrm{CT}.
    \label{eq:n-ct4}
\end{split}
\end{align}
\end{subequations}

At steady state, Eqs.~\ref{eq:ndot4}--\ref{eq:n-ct4} equal zero, and we get

\begin{equation}
    k_\mathrm{m} (n_0 + \Delta n) \Delta p = G_{h \nu} + k_\mathrm{s} n_\mathrm{{CT}},
\end{equation}

\begin{equation}
    (n_0 + \Delta n) \Delta p = \frac{G_{h \nu}}{\gamma k_\mathrm{L}}.
\end{equation}

If $\Delta n << n_0$,

\begin{equation}
    \Delta p = \frac{1}{n_0}\frac{G_{h \nu}}{\gamma k_\mathrm{L}}.
\end{equation}

In computing conductivity, let us neglect n. This assumption gives,

\begin{equation}
    \sigma = q \mu_n n_0 + q \mu_p \frac{G_{h \nu}}{n_0 \gamma k_\mathrm{{L}}}.
\end{equation}

In this case, conductivity is directly proportional to light intensity, which is what we see in our data.

\section{Error Estimation}

The error bars for the Langevin reduction factor were estimated as follows,

\begin{equation}
    \left(\frac{\sigma_{\gamma}}{\gamma}\right) ^2 = 
    \left(\frac{\sigma_{m}}{m}\right)^2 +
    \left(\frac{\sigma_{c}}{c}\right) ^2 +
    2 \left( \frac{\sigma_{m}}{m} \right) \left(\frac{\sigma_{c}}{c}\right) \rho_{mc}
\end{equation}
 
\noindent where $\gamma$ is the Langevin reduction factor; $m$ and $c$ are the slope and intercept for the linear fit of conductivity versus light intensity; $\sigma_{\gamma}$, $\sigma_m$, and $\sigma_c$ are the error for the Langevin reduction factor, slope, and intercept, respectively; and $\rho_{mc}$ is the correlation coefficient. 

\clearpage
\section{Broadband Local Dielectric Spectroscopy}

\begin{figure*}[h!]
    \begin{adjustwidth}{-1in}{-1in}
    \centering
    \includegraphics[width=6.6in]{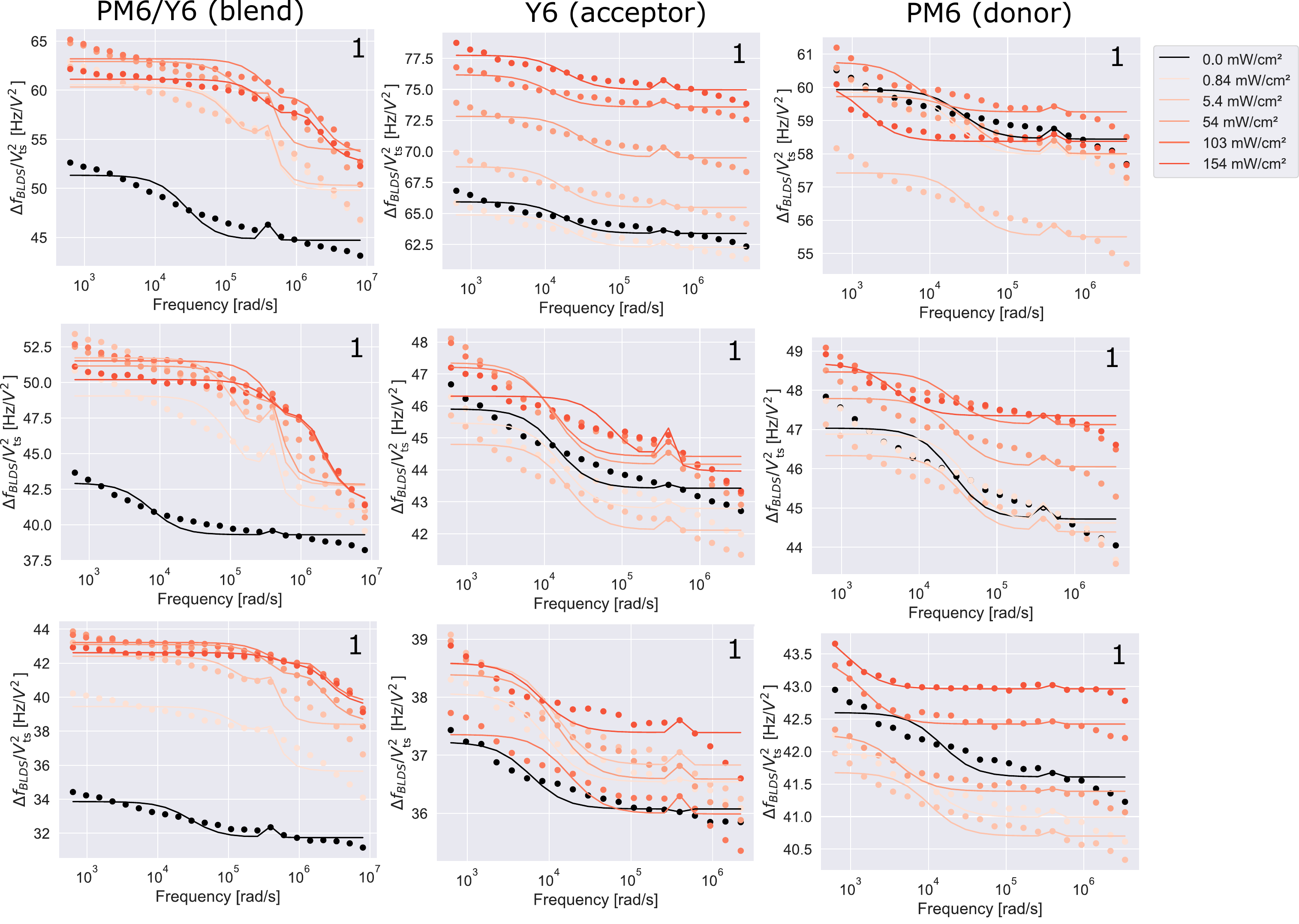}
    \end{adjustwidth}
    \caption[All BLDS spectra collected for samples containing the hole transport layer, PEDOT:PSS.]{All BLDS spectra collected for samples containing the hole transport layer, PEDOT:PSS. Numbers in the upper right corner of each spectra correspond to the cantilever number that was used to collect those data.}
    \label{fig:blds-pedot-pss}
\end{figure*}

\begin{figure*}[h!]
    \begin{adjustwidth}{-1in}{-1in}
    \centering
    \includegraphics[width=6.6in]{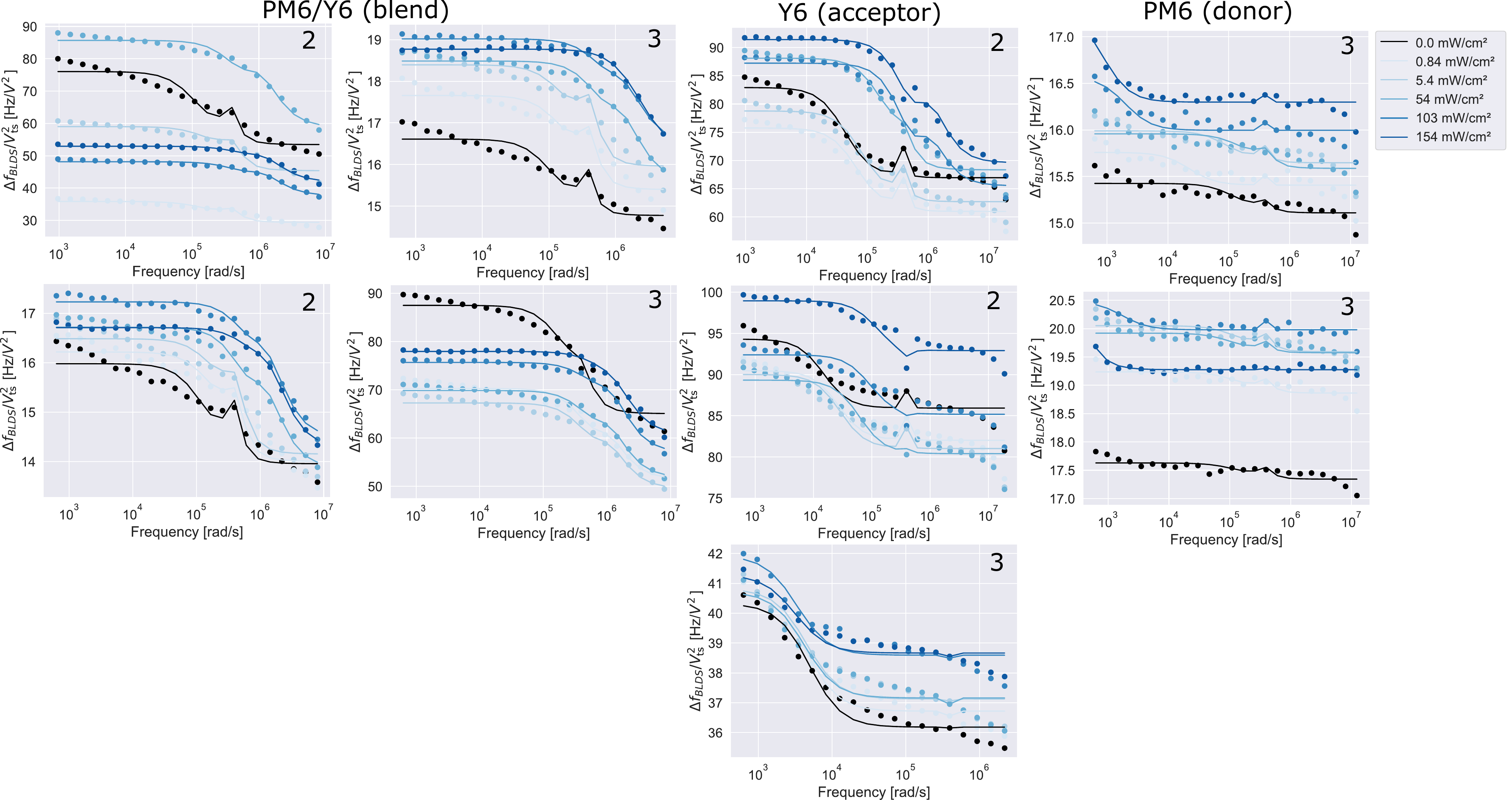}
    \end{adjustwidth}
    \caption[All BLDS spectra collected for samples without the hole transport layer.]{All BLDS spectra collected for samples without the hole transport layer. Numbers in the upper right corner of each spectra correspond to the cantilever number that was used to collect those data.}
    \label{fig:blds-ito}
\end{figure*}

Below are all of the BLDS spectra collected for the PM6:Y6 blend and controls, Y6 and PM6, both with (Fig.~\ref{fig:blds-pedot-pss}) and without (Fig.~\ref{fig:blds-ito}) the hole transport layer, PEDOT:PSS. The cantilever details for each spectra are listed in Table~\ref{tab:cant}. The BLDS experiment is described by Tirmzi \textit{et al.} \cite{Tirmzi2019feb}, with the amplitude modulation frequency $f_\mathrm{AM}$ = 20 Hz and the peak-to-peak voltage applied $V_\mathrm{ts}$ = 2 V using a waveform generator (Keysight 33622A). 

\begin{table}[h!]
    \centering
    \begin{tabular}{c@{\hspace{3ex}}c@{\hspace{3ex}}c@{\hspace{3ex}}c} \toprule
     Cantilever Number & $f_0$ [kHz] & $A_{pp}$ [nm] & Q  \\
     \midrule
     1 & 60.437 & 226.27 & 20202 \\
     2 & 63.670 & 157.44 & 19999 \\
     3 & 62.149 & 135.41 & 27778 \\
     \bottomrule
    \end{tabular}
    \caption{Details of all cantilevers (MikroMasch HQ:NSC18/Pt) used for the described experiments.}
    \label{tab:cant}
\end{table}

\FloatBarrier

\begin{figure}
    \centering
    \includegraphics[width=3.3in]{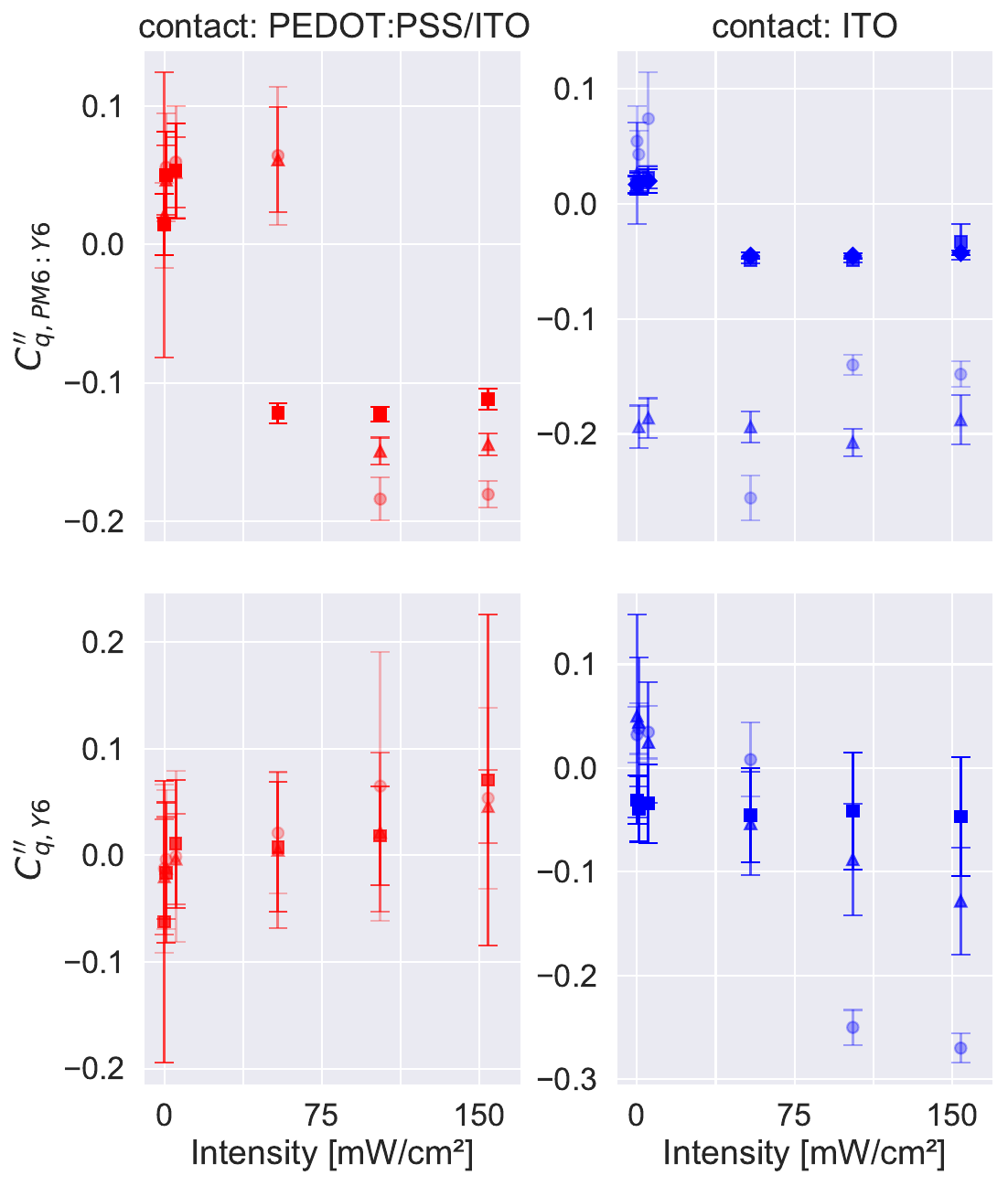}
    \caption[$C_q^{''}$ vs. light intensity for all samples obtained from the BLDS fits.]{$C_q^{''}$ vs. light intensity for all samples obtained from the BLDS fits. Each symbol corresponds to a different physical sample. Red data points are samples that contain PEDOT:PSS, and blue data points are samples without PEDOT:PSS. Data were collected in triplicate. All error bars are 1$\sigma$.}
    \label{fig:c1-vs-light}
\end{figure}

\begin{figure}
    \centering
    \includegraphics[width=3.3in]{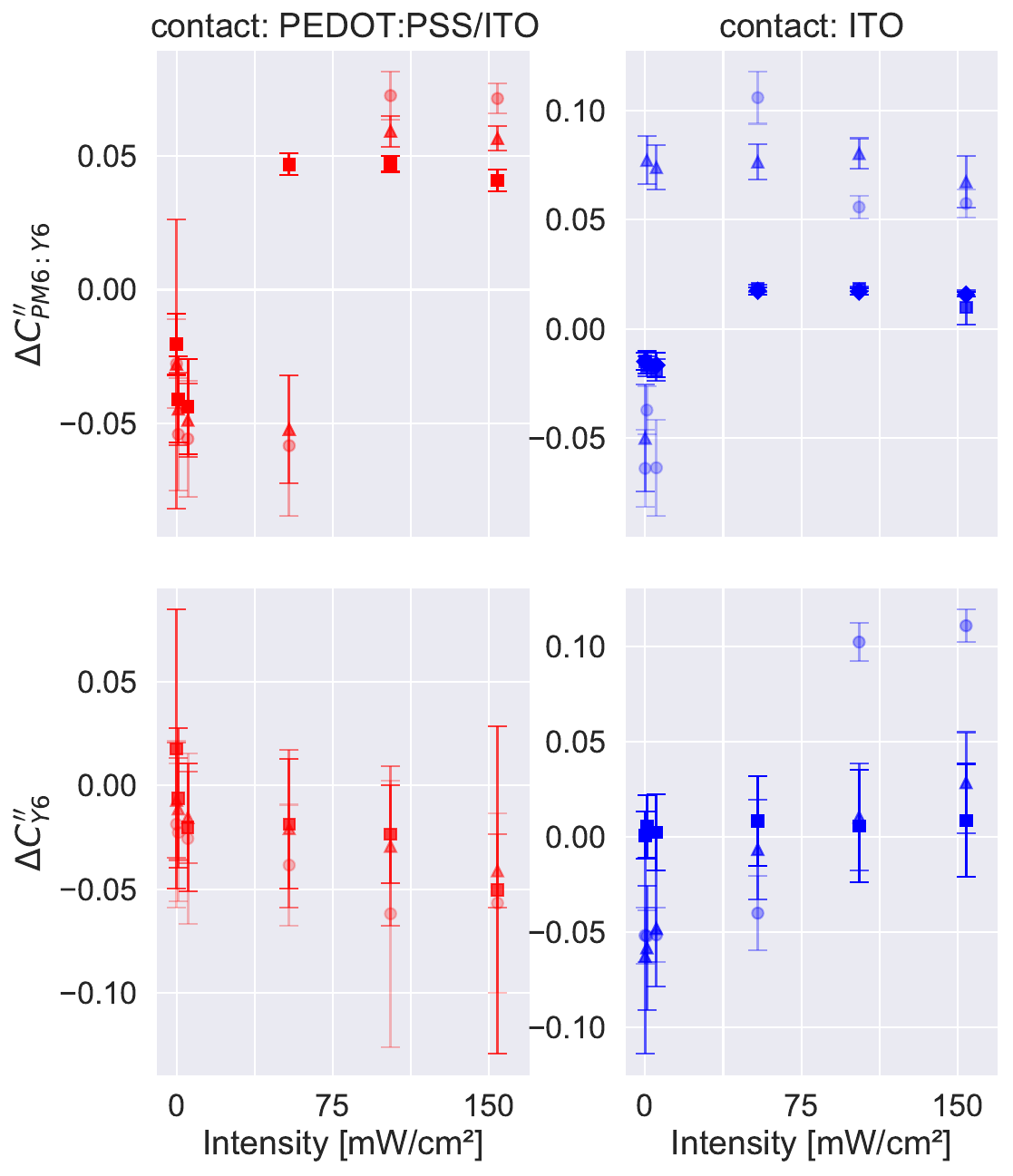}
    \caption[$\Delta C^{''}$ vs. light intensity for all samples obtained from the BLDS fits.]{$\Delta C^{''}$ vs. light intensity for all samples obtained from the BLDS fits. Each symbol corresponds to a different physical sample. Red data points are samples that contain PEDOT:PSS, and blue data points are samples without PEDOT:PSS. Data were collected in triplicate. All error bars are 1$\sigma$.}
    \label{fig:c2-vs-light}
\end{figure}

\begin{figure}
    \centering
    \includegraphics[width=3.3in]{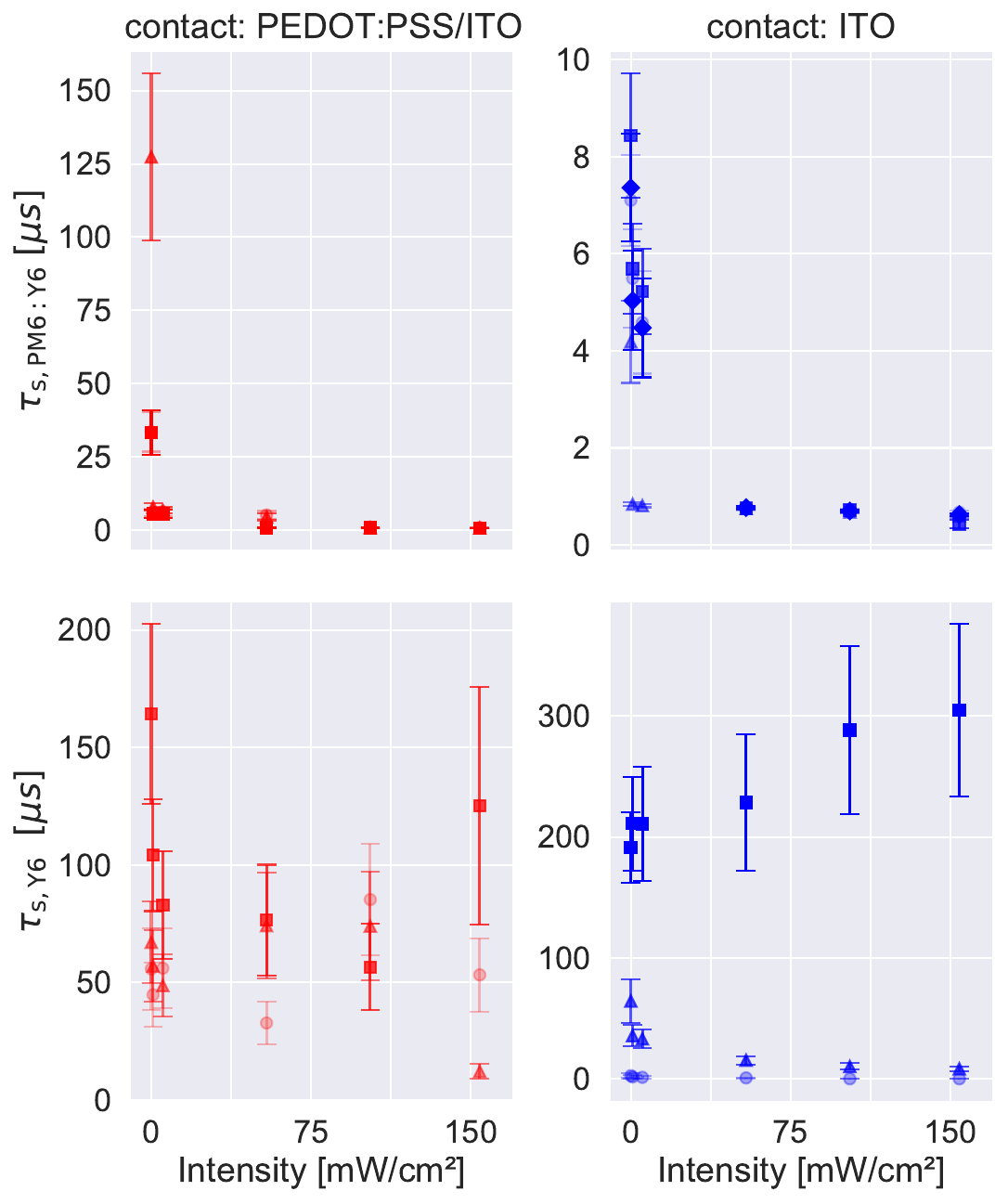}
    \caption[$\tau_\mathrm{s}$ vs. light intensity for all samples obtained from the BLDS fits.]{$\tau_\mathrm{s}$ vs. light intensity for all samples obtained from the BLDS fits. Each symbol corresponds to a different physical sample. Red data points are samples that contain PEDOT:PSS, and blue data points are samples without PEDOT:PSS. Data were collected in triplicate. All error bars are 1$\sigma$.}
    \label{fig:taus-vs-light}
\end{figure}

\begin{figure}
    \centering
    \includegraphics[width=3.3in]{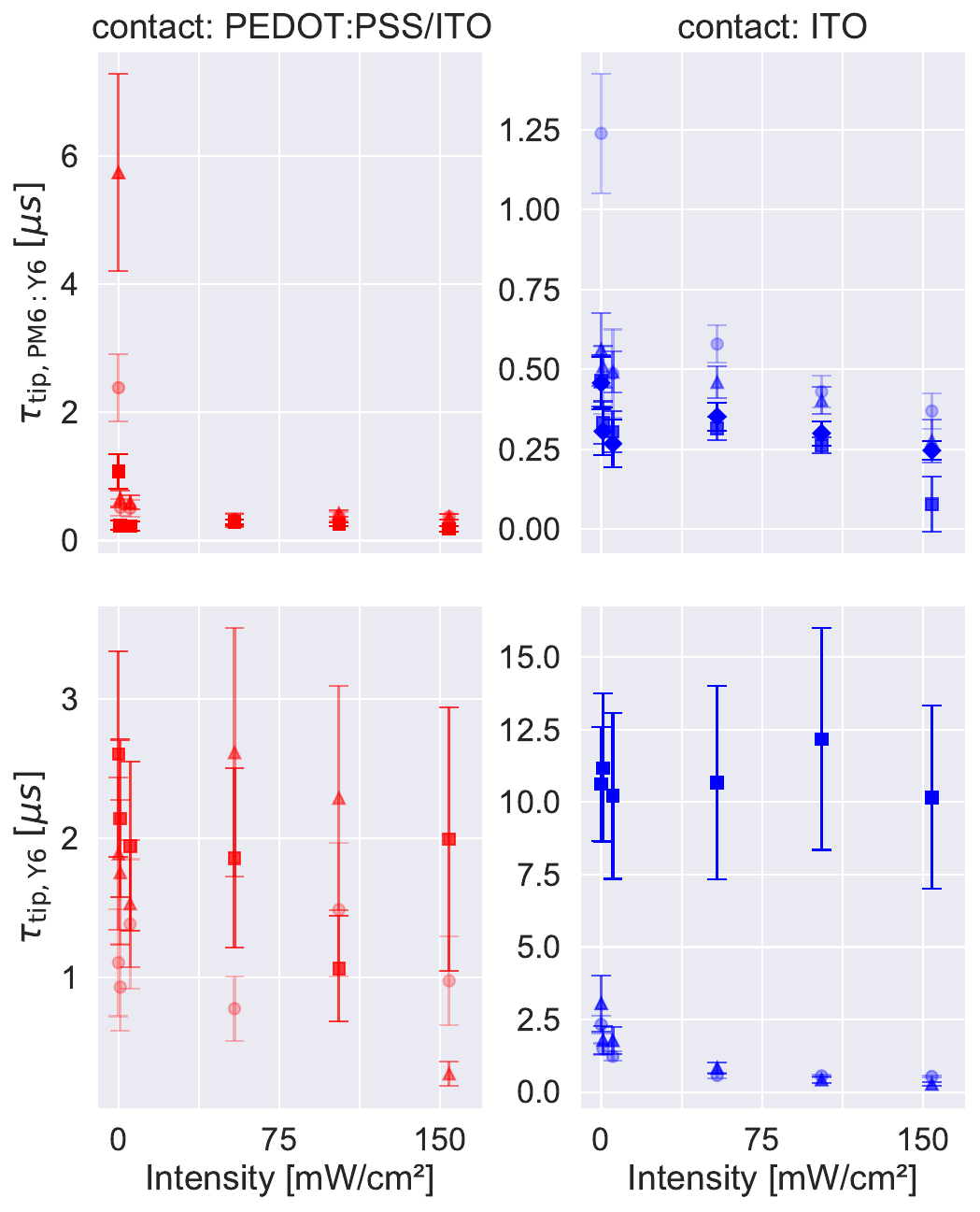}
    \caption[$\tau_\mathrm{tip}$ vs. light intensity for all samples obtained from the BLDS fits.]{$\tau_\mathrm{tip}$ vs. light intensity for all samples obtained from the BLDS fits. Each symbol corresponds to a different physical sample. Red data points are samples that contain PEDOT:PSS, and blue data points are samples without PEDOT:PSS. Data were collected in triplicate. All error bars are 1$\sigma$.}
    \label{fig:tautip-vs-light}
\end{figure}

Figs.~\ref{fig:c1-vs-light}--\ref{fig:tautip-vs-light} show BLDS fit parameters $C_q^{''}$, $\Delta C^{''}$, $\tau_\mathrm{s}$, and $\tau_\mathrm{tip}$ \emph{versus} light intensity. Tables~\ref{tab:fit-params-1}--\ref{tab:fit-params-13} list the fit parameters as well.

\FloatBarrier

\begin{table*}
    \begin{adjustwidth}{-1in}{-1in}
    \centering
    \begin{tabular}{c@{\hspace{3ex}}c@{\hspace{3ex}}c@{\hspace{3ex}}c@{\hspace{3ex}}c@{\hspace{3ex}}c@{\hspace{3ex}}c} \toprule
    Light Intensity & $C_q^{''}$ & $\Delta C^{''}$  & $\tau_s$    & $\tau_\mathrm{tip}$ \\
     \midrule
0 mW/cm$^2$    & 5.45e-02 $\pm$ 3.06e-02  & -6.38e-02 $\pm$ 1.77e-02 & 7.10e-06 $\pm$ 9.41e-07 & 1.24e-06 $\pm$ 1.87e-07 \\
0.84 mW/cm$^2$ & 4.30e-02 $\pm$ 2.06e-02  & -3.71e-02 $\pm$ 1.10e-02 & 5.49e-06 $\pm$ 1.01e-06 & 4.66e-07 $\pm$ 1.05e-07 \\
5.4 mW/cm$^2$ & 7.40e-02 $\pm$ 4.05e-02  & -6.35e-02 $\pm$ 2.21e-02 & 4.58e-06 $\pm$ 1.05e-06 & 4.87e-07 $\pm$ 1.38e-07 \\
54 mW/cm$^2$  & -2.56e-01 $\pm$ 1.95e-02 & 1.06e-01 $\pm$ 1.19e-02  & 7.47e-07 $\pm$ 2.72e-08 & 5.79e-07 $\pm$ 5.88e-08 \\
103 mW/cm$^2$ & -1.40e-01 $\pm$ 8.95e-03 & 5.58e-02 $\pm$ 5.28e-03  & 7.31e-07 $\pm$ 2.52e-08 & 4.31e-07 $\pm$ 4.89e-08 \\
154 mW/cm$^2$ & -1.48e-01 $\pm$ 1.12e-02 & 5.74e-02 $\pm$ 6.46e-03  & 6.75e-07 $\pm$ 3.27e-08 & 3.70e-07 $\pm$ 5.57e-08 \\
        \bottomrule
    \end{tabular}
    \end{adjustwidth}
    \caption{Fit parameters for PM6:Y6/ITO, $\circ$ dataset.}
    \label{tab:fit-params-1}
\end{table*}

\begin{table*}
    \begin{adjustwidth}{-1in}{-1in}
    \centering
    \begin{tabular}{c@{\hspace{3ex}}c@{\hspace{3ex}}c@{\hspace{3ex}}c@{\hspace{3ex}}c@{\hspace{3ex}}c@{\hspace{3ex}}c} \toprule
    Light Intensity & $C_q^{''}$ & $\Delta C^{''}$  & $\tau_s$    & $\tau_\mathrm{tip}$ \\
     \midrule
0 mW/cm$^2$    & 2.65e-02 $\pm$ 4.40e-02  & -5.00e-02 $\pm$ 2.44e-02 & 4.19e-06 $\pm$ 8.46e-07 & 5.63e-07 $\pm$ 1.14e-07 \\
0.84 mW/cm$^2$ & -1.94e-01 $\pm$ 1.84e-02 & 7.72e-02 $\pm$ 1.10e-02  & 8.42e-07 $\pm$ 4.09e-08 & 5.09e-07 $\pm$ 6.62e-08 \\
5.4 mW/cm$^2$ & -1.86e-01 $\pm$ 1.71e-02 & 7.39e-02 $\pm$ 1.02e-02  & 8.12e-07 $\pm$ 3.61e-08 & 4.92e-07 $\pm$ 6.49e-08 \\
54 mW/cm$^2$  & -1.94e-01 $\pm$ 1.36e-02 & 7.64e-02 $\pm$ 8.03e-03  & 7.54e-07 $\pm$ 2.44e-08 & 4.60e-07 $\pm$ 5.05e-08 \\
103 mW/cm$^2$ & -2.07e-01 $\pm$ 1.20e-02 & 8.03e-02 $\pm$ 6.97e-03  & 6.81e-07 $\pm$ 2.11e-08 & 4.04e-07 $\pm$ 4.16e-08 \\
154 mW/cm$^2$ & -1.88e-01 $\pm$ 2.13e-02 & 6.73e-02 $\pm$ 1.17e-02  & 6.00e-07 $\pm$ 4.65e-08 & 2.77e-07 $\pm$ 6.65e-08 \\

        \bottomrule
    \end{tabular}
    \end{adjustwidth}
    \caption{Fit parameters for PM6:Y6/ITO, $\triangle$ dataset.}
\end{table*}

\begin{table*}
    \begin{adjustwidth}{-1in}{-1in}
    \centering
    \begin{tabular}{c@{\hspace{3ex}}c@{\hspace{3ex}}c@{\hspace{3ex}}c@{\hspace{3ex}}c@{\hspace{3ex}}c@{\hspace{3ex}}c} \toprule
    Light Intensity & $C_q^{''}$ & $\Delta C^{''}$  & $\tau_s$    & $\tau_\mathrm{tip}$ \\
     \midrule
0 mW/cm$^2$    & 1.60e-02 $\pm$ 7.43e-03  & -1.47e-02 $\pm$ 3.90e-03 & 8.44e-06 $\pm$ 1.28e-06 & 4.64e-07 $\pm$ 8.10e-08 \\
0.84 mW/cm$^2$ & 1.97e-02 $\pm$ 8.78e-03  & -1.70e-02 $\pm$ 4.61e-03 & 5.70e-06 $\pm$ 9.32e-07 & 3.34e-07 $\pm$ 6.61e-08 \\
5.4 mW/cm$^2$ & 2.29e-02 $\pm$ 9.47e-03  & -1.89e-02 $\pm$ 4.97e-03 & 5.22e-06 $\pm$ 8.85e-07 & 3.06e-07 $\pm$ 6.37e-08 \\
54 mW/cm$^2$  & -4.88e-02 $\pm$ 2.68e-03 & 1.86e-02 $\pm$ 1.52e-03  & 7.53e-07 $\pm$ 2.33e-08 & 3.15e-07 $\pm$ 3.72e-08 \\
103 mW/cm$^2$ & -4.91e-02 $\pm$ 1.92e-03 & 1.83e-02 $\pm$ 1.07e-03  & 7.16e-07 $\pm$ 1.81e-08 & 2.62e-07 $\pm$ 2.55e-08 \\
154 mW/cm$^2$ & -3.31e-02 $\pm$ 1.57e-02 & 9.83e-03 $\pm$ 8.00e-03  & 4.33e-07 $\pm$ 8.89e-08 & 7.80e-08 $\pm$ 8.58e-08\\
    \bottomrule
    \end{tabular}
    \end{adjustwidth}
    \caption{Fit parameters for PM6:Y6/ITO, $\square$ dataset.}
\end{table*}

\begin{table*}
    \begin{adjustwidth}{-1in}{-1in}
    \centering
    \begin{tabular}{c@{\hspace{3ex}}c@{\hspace{3ex}}c@{\hspace{3ex}}c@{\hspace{3ex}}c@{\hspace{3ex}}c@{\hspace{3ex}}c} \toprule
    Light Intensity & $C_q^{''}$ & $\Delta C^{''}$  & $\tau_s$    & $\tau_\mathrm{tip}$ \\
     \midrule
0 mW/cm$^2$    & 1.68e-02 $\pm$ 7.41e-03  & -1.49e-02 $\pm$ 3.91e-03 & 7.36e-06 $\pm$ 1.11e-06 & 4.58e-07 $\pm$ 8.19e-08 \\
0.84 mW/cm$^2$ & 1.75e-02 $\pm$ 9.64e-03  & -1.53e-02 $\pm$ 5.07e-03 & 5.03e-06 $\pm$ 1.02e-06 & 3.07e-07 $\pm$ 7.52e-08 \\
5.4 mW/cm$^2$ & 1.98e-02 $\pm$ 1.07e-02  & -1.66e-02 $\pm$ 5.60e-03 & 4.48e-06 $\pm$ 1.02e-06 & 2.68e-07 $\pm$ 7.43e-08 \\
54 mW/cm$^2$  & -4.52e-02 $\pm$ 2.86e-03 & 1.75e-02 $\pm$ 1.65e-03  & 7.71e-07 $\pm$ 2.63e-08 & 3.53e-07 $\pm$ 4.39e-08 \\
103 mW/cm$^2$ & -4.53e-02 $\pm$ 2.58e-03 & 1.71e-02 $\pm$ 1.46e-03  & 7.01e-07 $\pm$ 2.54e-08 & 3.00e-07 $\pm$ 3.83e-08 \\
154 mW/cm$^2$ & -4.26e-02 $\pm$ 1.93e-03 & 1.57e-02 $\pm$ 1.07e-03  & 6.28e-07 $\pm$ 2.52e-08 & 2.47e-07 $\pm$ 2.90e-08 \\ 
    \bottomrule
    \end{tabular}
    \end{adjustwidth}
    \caption{Fit parameters for PM6:Y6/ITO, $\diamondsuit$ dataset.}
\end{table*}

\begin{table*}
    \begin{adjustwidth}{-1in}{-1in}
    \centering
    \begin{tabular}{c@{\hspace{3ex}}c@{\hspace{3ex}}c@{\hspace{3ex}}c@{\hspace{3ex}}c@{\hspace{3ex}}c@{\hspace{3ex}}c} \toprule
    Light Intensity & $C_q^{''}$ & $\Delta C^{''}$  & $\tau_s$    & $\tau_\mathrm{tip}$ \\
     \midrule
0 mW/cm$^2$    & 3.21e-02 $\pm$ 2.67e-02  & -5.17e-02 $\pm$ 1.48e-02 & 2.58e-06 $\pm$ 2.05e-06 & 2.33e-06 $\pm$ 3.15e-07 \\
0.84 mW/cm$^2$ & 3.82e-02 $\pm$ 2.46e-02  & -5.20e-02 $\pm$ 1.36e-02 & 1.61e-06 $\pm$ 1.48e-06 & 1.48e-06 $\pm$ 1.98e-07 \\
5.4 mW/cm$^2$ & 3.47e-02 $\pm$ 2.54e-02  & -5.14e-02 $\pm$ 1.41e-02 & 1.24e-06 $\pm$ 1.15e-06 & 1.23e-06 $\pm$ 1.61e-07 \\
54 mW/cm$^2$  & 8.20e-03 $\pm$ 3.54e-02  & -4.00e-02 $\pm$ 1.95e-02 & 7.81e-07 $\pm$ 3.25e-07 & 5.74e-07 $\pm$ 9.57e-08 \\
103 mW/cm$^2$ & -2.50e-01 $\pm$ 1.68e-02 & 1.02e-01 $\pm$ 1.01e-02  & 3.92e-08 $\pm$ 1.72e-07 & 5.54e-07 $\pm$ 4.66e-08 \\
154 mW/cm$^2$ & -2.70e-01 $\pm$ 1.42e-02 & 1.11e-01 $\pm$ 8.58e-03  & 2.53e-08 $\pm$ 1.24e-07 & 5.33e-07 $\pm$ 3.85e-08 \\ 
    \bottomrule
    \end{tabular}
    \end{adjustwidth}
    \caption{Fit parameters for Y6/ITO, $\circ$ dataset.}
\end{table*}

\begin{table*}
    \begin{adjustwidth}{-1in}{-1in}
    \centering
    \begin{tabular}{c@{\hspace{3ex}}c@{\hspace{3ex}}c@{\hspace{3ex}}c@{\hspace{3ex}}c@{\hspace{3ex}}c@{\hspace{3ex}}c} \toprule
    Light Intensity & $C_q^{''}$ & $\Delta C^{''}$  & $\tau_s$    & $\tau_\mathrm{tip}$ \\
     \midrule
0 mW/cm$^2$    & 5.02e-02 $\pm$ 9.78e-02  & -6.26e-02 $\pm$ 5.12e-02 & 6.43e-05 $\pm$ 1.80e-05 & 3.05e-06 $\pm$ 9.74e-07 \\
0.84 mW/cm$^2$ & 4.42e-02 $\pm$ 6.20e-02  & -5.81e-02 $\pm$ 3.26e-02 & 3.57e-05 $\pm$ 8.91e-06 & 1.79e-06 $\pm$ 4.93e-07 \\
5.4 mW/cm$^2$ & 2.49e-02 $\pm$ 5.82e-02  & -4.80e-02 $\pm$ 3.07e-02 & 3.31e-05 $\pm$ 7.93e-06 & 1.78e-06 $\pm$ 4.69e-07 \\
54 mW/cm$^2$  & -5.34e-02 $\pm$ 4.98e-02 & -6.47e-03 $\pm$ 2.62e-02 & 1.54e-05 $\pm$ 3.49e-06 & 8.30e-07 $\pm$ 2.01e-07 \\
103 mW/cm$^2$ & -8.81e-02 $\pm$ 5.40e-02 & 1.05e-02 $\pm$ 2.81e-02  & 1.02e-05 $\pm$ 2.67e-06 & 4.29e-07 $\pm$ 1.16e-07 \\
154 mW/cm$^2$ & -1.28e-01 $\pm$ 5.14e-02 & 2.85e-02 $\pm$ 2.65e-02  & 8.25e-06 $\pm$ 2.19e-06 & 2.73e-07 $\pm$ 7.09e-08\\
    \bottomrule
    \end{tabular}
    \end{adjustwidth}
    \caption{Fit parameters for Y6/ITO, $\triangle$ dataset.}
\end{table*}

\begin{table*}
    \begin{adjustwidth}{-1in}{-1in}
    \centering
    \begin{tabular}{c@{\hspace{3ex}}c@{\hspace{3ex}}c@{\hspace{3ex}}c@{\hspace{3ex}}c@{\hspace{3ex}}c@{\hspace{3ex}}c} \toprule
    Light Intensity & $C_q^{''}$ & $\Delta C^{''}$  & $\tau_s$    & $\tau_\mathrm{tip}$ \\
     \midrule
0 mW/cm$^2$    & -3.07e-02 $\pm$ 2.34e-02 & 8.66e-04 $\pm$ 1.23e-02 & 1.91e-04 $\pm$ 2.89e-05 & 1.06e-05 $\pm$ 1.98e-06 \\
0.84 mW/cm$^2$ & -3.97e-02 $\pm$ 3.13e-02 & 5.43e-03 $\pm$ 1.65e-02 & 2.11e-04 $\pm$ 3.87e-05 & 1.12e-05 $\pm$ 2.56e-06 \\
5.4 mW/cm$^2$ & -3.42e-02 $\pm$ 3.79e-02 & 2.51e-03 $\pm$ 1.98e-02 & 2.11e-04 $\pm$ 4.73e-05 & 1.02e-05 $\pm$ 2.86e-06 \\
54 mW/cm$^2$  & -4.54e-02 $\pm$ 4.53e-02 & 8.37e-03 $\pm$ 2.37e-02 & 2.28e-04 $\pm$ 5.65e-05 & 1.07e-05 $\pm$ 3.33e-06 \\
103 mW/cm$^2$ & -4.14e-02 $\pm$ 5.66e-02 & 5.82e-03 $\pm$ 2.95e-02 & 2.88e-04 $\pm$ 6.92e-05 & 1.22e-05 $\pm$ 3.83e-06 \\
154 mW/cm$^2$ & -4.66e-02 $\pm$ 5.74e-02 & 8.70e-03 $\pm$ 2.97e-02 & 3.05e-04 $\pm$ 7.15e-05 & 1.02e-05 $\pm$ 3.15e-06 \\
    \bottomrule
    \end{tabular}
    \end{adjustwidth}
    \caption{Fit parameters Y6/ITO, $\square$ dataset.}
\end{table*}

\begin{table*}
    \begin{adjustwidth}{-1in}{-1in}
    \centering
    \begin{tabular}{c@{\hspace{3ex}}c@{\hspace{3ex}}c@{\hspace{3ex}}c@{\hspace{3ex}}c@{\hspace{3ex}}c@{\hspace{3ex}}c} \toprule
    Light Intensity & $C_q^{''}$ & $\Delta C^{''}$  & $\tau_s$    & $\tau_\mathrm{tip}$ \\
     \midrule
0 mW/cm$^2$ &
  1.37e-02 $\pm$ 3.09e-02 &
  -2.77e-02 $\pm$ 1.66e-02 &
  3.35e-05 $\pm$ 6.76e-06 &
  2.39e-06 $\pm$ 5.22e-07 \\
0.84 mW/cm$^2$ &
  5.57e-02 $\pm$ 3.91e-02 &
  -5.40e-02 $\pm$ 2.10e-02 &
  5.92e-06 $\pm$ 1.25e-06 &
  5.18e-07 $\pm$ 1.29e-07 \\
5.4 mW/cm$^2$ &
  5.94e-02 $\pm$ 4.04e-02 &
  -5.57e-02 $\pm$ 2.16e-02 &
  6.21e-06 $\pm$ 1.34e-06 &
  5.05e-07 $\pm$ 1.29e-07 \\
54 mW/cm$^2$ &
  6.41e-02 $\pm$ 4.98e-02 &
  -5.82e-02 $\pm$ 2.63e-02 &
  5.13e-06 $\pm$ 1.38e-06 &
  3.31e-07 $\pm$ 1.06e-07 \\
103 mW/cm$^2$ &
  -1.84e-01 $\pm$ 1.55e-02 &
  7.25e-02 $\pm$ 8.99e-03 &
  8.18e-07 $\pm$ 3.98e-08 &
  3.80e-07 $\pm$ 6.34e-08 \\
154 mW/cm$^2$ &
  -1.81e-01 $\pm$ 9.77e-03 &
  7.14e-02 $\pm$ 5.69e-03 &
  8.15e-07 $\pm$ 2.69e-08 &
  3.71e-07 $\pm$ 4.18e-08
 \\
    \bottomrule
    \end{tabular}
    \end{adjustwidth}
    \caption{Fit parameters for PM6:Y6/PEDOT:PSS/ITO, $\circ$ dataset.}
\end{table*}

\begin{table*}
    \begin{adjustwidth}{-1in}{-1in}
    \centering
    \begin{tabular}{c@{\hspace{3ex}}c@{\hspace{3ex}}c@{\hspace{3ex}}c@{\hspace{3ex}}c@{\hspace{3ex}}c@{\hspace{3ex}}c} \toprule
    Light Intensity & $C_q^{''}$ & $\Delta C^{''}$  & $\tau_s$    & $\tau_\mathrm{tip}$ \\
     \midrule
0 mW/cm$^2$    & 2.13e-02 $\pm$ 1.03e-01 & -2.78e-02 $\pm$ 5.39e-02 & 1.27e-04 $\pm$ 2.86e-05 & 5.75e-06 $\pm$ 1.54e-06 \\
0.84 mW/cm$^2$ & 4.65e-02 $\pm$ 2.52e-02  & -4.46e-02 $\pm$ 1.36e-02 & 7.81e-06 $\pm$ 1.29e-06 & 6.51e-07 $\pm$ 1.25e-07 \\
5.4 mW/cm$^2$ & 5.22e-02 $\pm$ 2.55e-02  & -4.87e-02 $\pm$ 1.37e-02 & 6.80e-06 $\pm$ 1.06e-06 & 5.93e-07 $\pm$ 1.09e-07 \\
54 mW/cm$^2$  & 6.11e-02 $\pm$ 3.78e-02  & -5.21e-02 $\pm$ 2.00e-02 & 4.49e-06 $\pm$ 1.16e-06 & 3.14e-07 $\pm$ 9.71e-08 \\
103 mW/cm$^2$ & -1.49e-01 $\pm$ 9.80e-03 & 5.92e-02 $\pm$ 5.76e-03  & 8.10e-07 $\pm$ 2.83e-08 & 4.24e-07 $\pm$ 4.91e-08 \\
154 mW/cm$^2$ & -1.44e-01 $\pm$ 7.91e-03 & 5.65e-02 $\pm$ 4.58e-03  & 7.39e-07 $\pm$ 2.47e-08 & 3.69e-07 $\pm$ 4.14e-08  \\
    \bottomrule
    \end{tabular}
    \end{adjustwidth}
    \caption{Fit parameters for PM6:Y6/PEDOT:PSS/ITO, $\triangle$ dataset.}
\end{table*}

\begin{table*}
    \begin{adjustwidth}{-1in}{-1in}
    \centering
    \begin{tabular}{c@{\hspace{3ex}}c@{\hspace{3ex}}c@{\hspace{3ex}}c@{\hspace{3ex}}c@{\hspace{3ex}}c@{\hspace{3ex}}c} \toprule
    Light Intensity & $C_q^{''}$ & $\Delta C^{''}$  & $\tau_s$    & $\tau_\mathrm{tip}$ \\
     \midrule
0 mW/cm$^2$    & 1.44e-02 $\pm$ 2.21e-02  & -2.04e-02 $\pm$ 1.14e-02 & 3.32e-05 $\pm$ 7.54e-06 & 1.08e-06 $\pm$ 2.68e-07 \\
0.84 mW/cm$^2$ & 5.01e-02 $\pm$ 3.11e-02  & -4.10e-02 $\pm$ 1.61e-02 & 5.63e-06 $\pm$ 1.42e-06 & 2.37e-07 $\pm$ 7.37e-08 \\
5.4 mW/cm$^2$ & 5.31e-02 $\pm$ 3.42e-02  & -4.37e-02 $\pm$ 1.77e-02 & 5.50e-06 $\pm$ 1.43e-06 & 2.25e-07 $\pm$ 7.24e-08 \\
54 mW/cm$^2$  & -1.22e-01 $\pm$ 7.22e-03 & 4.68e-02 $\pm$ 4.10e-03  & 7.90e-07 $\pm$ 3.23e-08 & 2.92e-07 $\pm$ 4.36e-08 \\
103 mW/cm$^2$ & -1.23e-01 $\pm$ 5.31e-03 & 4.70e-02 $\pm$ 3.00e-03  & 7.62e-07 $\pm$ 2.80e-08 & 2.59e-07 $\pm$ 3.25e-08 \\
154 mW/cm$^2$ & -1.12e-01 $\pm$ 7.49e-03 & 4.08e-02 $\pm$ 4.08e-03  & 6.48e-07 $\pm$ 5.44e-08 & 1.81e-07 $\pm$ 4.36e-08 \\
    \bottomrule
    \end{tabular}
    \end{adjustwidth}
    \caption{Fit parameters for PM6:Y6/PEDOT:PSS/ITO, $\square$ dataset.}
\end{table*}

\begin{table*}
    \begin{adjustwidth}{-1in}{-1in}
    \centering
    \begin{tabular}{c@{\hspace{3ex}}c@{\hspace{3ex}}c@{\hspace{3ex}}c@{\hspace{3ex}}c@{\hspace{3ex}}c@{\hspace{3ex}}c} \toprule
    Light Intensity & $C_q^{''}$ & $\Delta C^{''}$  & $\tau_s$    & $\tau_\mathrm{tip}$ \\
     \midrule
0 mW/cm$^2$    & -1.24e-02 $\pm$ 7.86e-02 & -1.86e-02 $\pm$ 4.01e-02 & 5.57e-05 $\pm$ 1.74e-05 & 1.11e-06 $\pm$ 3.86e-07 \\
0.84 mW/cm$^2$ & -3.76e-03 $\pm$ 6.52e-02 & -2.26e-02 $\pm$ 3.33e-02 & 4.48e-05 $\pm$ 1.37e-05 & 9.29e-07 $\pm$ 3.14e-07 \\
5.4 mW/cm$^2$ & -1.15e-03 $\pm$ 8.00e-02 & -2.55e-02 $\pm$ 4.10e-02 & 5.60e-05 $\pm$ 1.69e-05 & 1.38e-06 $\pm$ 4.66e-07 \\
54 mW/cm$^2$  & 2.09e-02 $\pm$ 5.69e-02  & -3.83e-02 $\pm$ 2.91e-02 & 3.27e-05 $\pm$ 9.05e-06 & 7.74e-07 $\pm$ 2.32e-07 \\
103 mW/cm$^2$ & 6.49e-02 $\pm$ 1.26e-01  & -6.17e-02 $\pm$ 6.41e-02 & 8.53e-05 $\pm$ 2.38e-05 & 1.49e-06 $\pm$ 4.79e-07 \\
154 mW/cm$^2$ & 5.36e-02 $\pm$ 8.47e-02  & -5.66e-02 $\pm$ 4.31e-02 & 5.32e-05 $\pm$ 1.56e-05 & 9.75e-07 $\pm$ 3.20e-07 \\
    \bottomrule
    \end{tabular}
    \end{adjustwidth}
    \caption{Fit parameters for Y6/PEDOT:PSS/ITO, $\circ$ dataset.}
\end{table*}

\begin{table*}
    \begin{adjustwidth}{-1in}{-1in}
    \centering
    \begin{tabular}{c@{\hspace{3ex}}c@{\hspace{3ex}}c@{\hspace{3ex}}c@{\hspace{3ex}}c@{\hspace{3ex}}c@{\hspace{3ex}}c} \toprule
    Light Intensity & $C_q^{''}$ & $\Delta C^{''}$  & $\tau_s$    & $\tau_\mathrm{tip}$ \\
     \midrule
0 mW/cm$^2$    & -2.02e-02 $\pm$ 5.41e-02 & -7.10e-03 $\pm$ 2.78e-02 & 6.71e-05 $\pm$ 1.73e-05 & 1.89e-06 $\pm$ 5.48e-07 \\
0.84 mW/cm$^2$ & -1.19e-02 $\pm$ 4.78e-02 & -1.12e-02 $\pm$ 2.46e-02 & 5.70e-05 $\pm$ 1.52e-05 & 1.75e-06 $\pm$ 5.20e-07 \\
5.4 mW/cm$^2$ & -3.43e-03 $\pm$ 4.25e-02 & -1.54e-02 $\pm$ 2.19e-02 & 4.87e-05 $\pm$ 1.33e-05 & 1.53e-06 $\pm$ 4.58e-07 \\
54 mW/cm$^2$  & 5.05e-03 $\pm$ 7.30e-02  & -2.08e-02 $\pm$ 3.78e-02 & 7.42e-05 $\pm$ 2.24e-05 & 2.62e-06 $\pm$ 8.95e-07 \\
103 mW/cm$^2$ & 2.17e-02 $\pm$ 7.44e-02  & -2.92e-02 $\pm$ 3.84e-02 & 7.40e-05 $\pm$ 2.30e-05 & 2.29e-06 $\pm$ 8.05e-07 \\
154 mW/cm$^2$ & 4.58e-02 $\pm$ 3.44e-02  & -4.10e-02 $\pm$ 1.76e-02 & 1.22e-05 $\pm$ 3.26e-06 & 3.05e-07 $\pm$ 8.84e-08 \\
    \bottomrule
    \end{tabular}
    \end{adjustwidth}
    \caption{Fit parameters for Y6/PEDOT:PSS/ITO, $\triangle$ dataset.}
\end{table*}

\begin{table*}[h!]
    \begin{adjustwidth}{-1in}{-1in}
    \centering
    \begin{tabular}{c@{\hspace{3ex}}c@{\hspace{3ex}}c@{\hspace{3ex}}c@{\hspace{3ex}}c@{\hspace{3ex}}c@{\hspace{3ex}}c} \toprule
    Light Intensity & $C_q^{''}$ & $\Delta C^{''}$  & $\tau_s$    & $\tau_\mathrm{tip}$ \\
     \midrule
0 mW/cm$^2$    & -6.23e-02 $\pm$ 1.32e-01 & 1.76e-02 $\pm$ 6.72e-02  & 1.64e-04 $\pm$ 3.83e-05 & 2.61e-06 $\pm$ 7.38e-07 \\
0.84 mW/cm$^2$ & -1.63e-02 $\pm$ 6.59e-02 & -6.09e-03 $\pm$ 3.36e-02 & 1.04e-04 $\pm$ 2.38e-05 & 2.14e-06 $\pm$ 5.68e-07 \\
5.4 mW/cm$^2$ & 1.10e-02 $\pm$ 6.01e-02  & -2.02e-02 $\pm$ 3.07e-02 & 8.30e-05 $\pm$ 2.28e-05 & 1.94e-06 $\pm$ 6.07e-07 \\
54 mW/cm$^2$  & 7.81e-03 $\pm$ 6.09e-02  & -1.86e-02 $\pm$ 3.12e-02 & 7.65e-05 $\pm$ 2.36e-05 & 1.86e-06 $\pm$ 6.45e-07 \\
103 mW/cm$^2$ & 1.82e-02 $\pm$ 4.64e-02  & -2.34e-02 $\pm$ 2.37e-02 & 5.65e-05 $\pm$ 1.84e-05 & 1.06e-06 $\pm$ 3.82e-07 \\
154 mW/cm$^2$ & 7.05e-02 $\pm$ 1.55e-01  & -5.04e-02 $\pm$ 7.89e-02 & 1.25e-04 $\pm$ 5.05e-05 & 1.99e-06 $\pm$ 9.49e-07 \\

    \bottomrule
    \end{tabular}
    \end{adjustwidth}
    \caption{Fit parameters for Y6/PEDOT:PSS/ITO, $\square$ dataset.}
    \label{tab:fit-params-13}
\end{table*}

\begin{table}[h!]
\begin{adjustwidth}{-1in}{-1in}
\centering
\begin{tabular}{c@{\hspace{3ex}}r@{\hspace{2ex}}r@{\hspace{3ex}}r@{\hspace{2ex}}r@{\hspace{3ex}}r@{\hspace{2ex}}r@{\hspace{3ex}}r@{\hspace{3ex}}r@{\hspace{3ex}}} \toprule
 \multicolumn{1}{r}{Dataset:} & \multicolumn{2}{c}{$\circ$} & \multicolumn{2}{c}{$\triangle$} & \multicolumn{2}{c}{$\square$} & \multicolumn{2}{c}{$\diamondsuit$} \\
 \cmidrule(rl){2-3} \cmidrule(rl){4-5} \cmidrule(rl){6-7}\cmidrule(rl){8-9}
     Light Intensity & $R_\mathrm{s}$ [$\Omega$]       & $C_\mathrm{s}$ [F]       & $R_\mathrm{s}$ [$\Omega$]       & $C_\mathrm{s}$ [F]     & $R_\mathrm{s}$ [$\Omega$]       & $C_\mathrm{s}$ [F]       & $R_\mathrm{s}$ [$\Omega$]       & $C_\mathrm{s}$ [F]       \\
     \midrule
0 mW/cm$^2$    & 2.55E+11 & 2.79E-17 & 1.16E+11 & 3.62E-17 & 9.56E+10 & 8.83E-17 & 9.42E+10 & 7.82E-17 \\
0.84 mW/cm$^2$ & 9.59E+10 & 5.73E-17 & 1.05E+11 & 8.04E-18 & 6.88E+10 & 8.28E-17 & 6.32E+10 & 7.97E-17 \\
5.4 mW/cm$^2$ & 1.00E+11 & 4.57E-17 & 1.01E+11 & 8.02E-18 & 6.29E+10 & 8.31E-17 & 5.52E+10 & 8.11E-17 \\
54 mW/cm$^2$  & 1.19E+11 & 6.27E-18 & 9.47E+10 & 7.96E-18 & 6.49E+10 & 1.16E-17 & 7.26E+10 & 1.06E-17 \\
103 mW/cm$^2$ & 8.86E+10 & 8.25E-18 & 8.31E+10 & 8.20E-18 & 5.40E+10 & 1.33E-17 & 6.18E+10 & 1.13E-17 \\
154 mW/cm$^2$ & 7.61E+10 & 8.87E-18 & 5.69E+10 & 1.05E-17 & 1.61E+10 & 2.70E-17 & 5.07E+10 & 1.24E-17
\\
\bottomrule
\end{tabular}
\end{adjustwidth}
\caption{$R_\mathrm{s}$ and $C_\mathrm{s}$ at each light intensity for all PM6:Y6/ITO samples.}
\label{tab:rs-cs-1}
\end{table}

\begin{table}[h!]
\begin{tabular}{c@{\hspace{3ex}}r@{\hspace{2ex}}r@{\hspace{3ex}}r@{\hspace{2ex}}r@{\hspace{3ex}}r@{\hspace{2ex}}r@{\hspace{3ex}}r@{\hspace{3ex}}r@{\hspace{3ex}}} \toprule
 \multicolumn{1}{r}{Dataset:} & \multicolumn{2}{c}{$\circ$} & \multicolumn{2}{c}{$\triangle$} & \multicolumn{2}{c}{$\square$}  \\
 \cmidrule(rl){2-3} \cmidrule(rl){4-5} \cmidrule(rl){6-7}
     Light Intensity & $R_\mathrm{s}$ [$\Omega$]       & $C_\mathrm{s}$ [F]       & $R_\mathrm{s}$ [$\Omega$]       & $C_\mathrm{s}$ [F]     & $R_\mathrm{s}$ [$\Omega$]       & $C_\mathrm{s}$ [F]           \\
     \midrule
0 mW/cm$^2$    & 4.79E+11 & 5.39E-18 & 6.28E+11 & 1.02E-16 & 2.19E+12 & 8.75E-17 \\
0.84 mW/cm$^2$ & 3.05E+11 & 5.29E-18 & 3.68E+11 & 9.72E-17 & 2.30E+12 & 9.18E-17 \\
5.4 mW/cm$^2$ & 2.53E+11 & 4.91E-18 & 3.66E+11 & 9.04E-17 & 2.10E+12 & 1.00E-16 \\
54 mW/cm$^2$  & 1.18E+11 & 6.61E-18 & 1.71E+11 & 9.02E-17 & 2.20E+12 & 1.04E-16 \\
103 mW/cm$^2$ & 1.14E+11 & 3.44E-19 & 8.82E+10 & 1.16E-16 & 2.51E+12 & 1.15E-16 \\
154 mW/cm$^2$ & 1.10E+11 & 2.31E-19 & 5.62E+10 & 1.47E-16 & 2.09E+12 & 1.46E-16
\\
     \bottomrule
\end{tabular}
\caption{$R_\mathrm{s}$ and $C_\mathrm{s}$ at each light intensity for all Y6/ITO samples.}
\label{rs-cs-2}
\end{table}

\begin{table}[h!]
\begin{tabular}{c@{\hspace{3ex}}r@{\hspace{2ex}}r@{\hspace{3ex}}r@{\hspace{2ex}}r@{\hspace{3ex}}r@{\hspace{2ex}}r@{\hspace{3ex}}r@{\hspace{3ex}}r@{\hspace{3ex}}} \toprule
 \multicolumn{1}{r}{Dataset:} & \multicolumn{2}{c}{$\circ$} & \multicolumn{2}{c}{$\triangle$} & \multicolumn{2}{c}{$\square$} \\
 \cmidrule(rl){2-3} \cmidrule(rl){4-5} \cmidrule(rl){6-7}
     Light Intensity & $R_\mathrm{s}$ [$\Omega$]       & $C_\mathrm{s}$ [F]       & $R_\mathrm{s}$ [$\Omega$]       & $C_\mathrm{s}$ [F]     & $R_\mathrm{s}$ [$\Omega$]       & $C_\mathrm{s}$ [F]         \\
     \midrule
0 mW/cm$^2$    & 4.91E+11 & 6.83E-17 & 1.18E+12 & 1.08E-16 & 2.23E+11 & 1.49E-16 \\
0.84 mW/cm$^2$ & 1.07E+11 & 5.56E-17 & 1.34E+11 & 5.83E-17 & 4.87E+10 & 1.16E-16 \\
5.4 mW/cm$^2$ & 1.04E+11 & 5.98E-17 & 1.22E+11 & 5.58E-17 & 4.63E+10 & 1.19E-16 \\
54 mW/cm$^2$  & 6.81E+10 & 7.52E-17 & 6.45E+10 & 6.96E-17 & 6.00E+10 & 1.32E-17 \\
103 mW/cm$^2$ & 7.82E+10 & 1.05E-17 & 8.72E+10 & 9.29E-18 & 5.32E+10 & 1.43E-17 \\
154 mW/cm$^2$ & 7.63E+10 & 1.07E-17 & 7.59E+10 & 9.73E-18 & 3.72E+10 & 1.74E-17
    \\
    \bottomrule
\end{tabular}
\caption{$R_\mathrm{s}$ and $C_\mathrm{s}$ at each light intensity for all PM6:Y6/PEDOT:PSS/ITO samples.}
\label{tab:rs-cs-3}
\end{table}

\begin{table}[h!]
\begin{tabular}{c@{\hspace{3ex}}r@{\hspace{2ex}}r@{\hspace{3ex}}r@{\hspace{2ex}}r@{\hspace{3ex}}r@{\hspace{2ex}}r@{\hspace{3ex}}r@{\hspace{3ex}}r@{\hspace{3ex}}} \toprule
 \multicolumn{1}{r}{Dataset:} & \multicolumn{2}{c}{$\circ$} & \multicolumn{2}{c}{$\triangle$} & \multicolumn{2}{c}{$\square$} \\
 \cmidrule(rl){2-3} \cmidrule(rl){4-5} \cmidrule(rl){6-7}
     Light Intensity & $R_\mathrm{s}$ [$\Omega$]       & $C_\mathrm{s}$ [F]       & $R_\mathrm{s}$ [$\Omega$]       & $C_\mathrm{s}$ [F]     & $R_\mathrm{s}$ [$\Omega$]       & $C_\mathrm{s}$ [F]        \\
     \midrule
0 mW/cm$^2$    & 2.27E+11 & 2.45E-16 & 3.89E+11 & 1.73E-16 & 5.36E+11 & 3.06E-16 \\
0.84 mW/cm$^2$ & 1.91E+11 & 2.34E-16 & 3.61E+11 & 1.58E-16 & 4.41E+11 & 2.36E-16 \\
5.4 mW/cm$^2$ & 2.84E+11 & 1.97E-16 & 3.15E+11 & 1.55E-16 & 4.00E+11 & 2.08E-16 \\
54 mW/cm$^2$  & 1.59E+11 & 2.06E-16 & 5.38E+11 & 1.38E-16 & 3.82E+11 & 2.00E-16 \\
103 mW/cm$^2$ & 3.06E+11 & 2.79E-16 & 4.71E+11 & 1.57E-16 & 2.18E+11 & 2.59E-16 \\
154 mW/cm$^2$ & 2.01E+11 & 2.65E-16 & 6.27E+10 & 1.94E-16 & 4.10E+11 & 3.05E-16
    \\
    \bottomrule
\end{tabular}
\caption{$R_\mathrm{s}$ and $C_\mathrm{s}$ at each light intensity for all Y6/PEDOT:PSS/ITO samples.}
\label{tab:rs-cs-4}
\end{table}

\clearpage
\section{Atomic Force Microscopy}

\begin{figure}
    \begin{adjustwidth}{-1in}{-1in}
    \centering
    \includegraphics[width=6.6in]{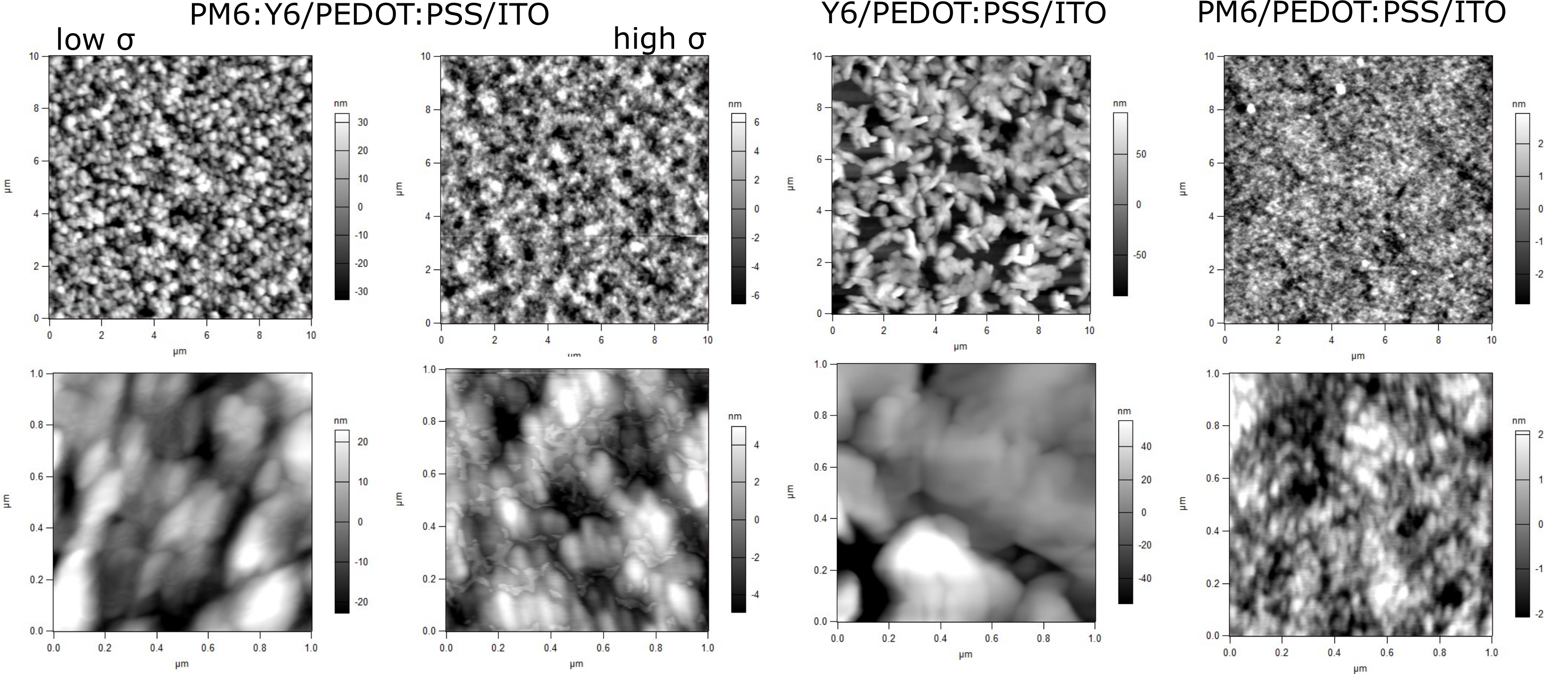}
    \end{adjustwidth}
    \caption[AFMs of PM6:Y6, Y6, and PM6 samples on PEDOT:PSS.]{AFMs of PM6:Y6, Y6, and PM6 samples on PEDOT:PSS. PM6:Y6/PEDOT:PSS/ITO ``low $\sigma$" was the second sample prepared ($\triangle$), with medium conductivity and ``high $\sigma$" was the third sample prepared ($\square$), with the highest conductivity of the 3 samples.}
    \label{fig:AFMs-pedot}
\end{figure}

\begin{figure}
    \begin{adjustwidth}{-1in}{-1in}
    \centering
    \includegraphics[width=6.6in]{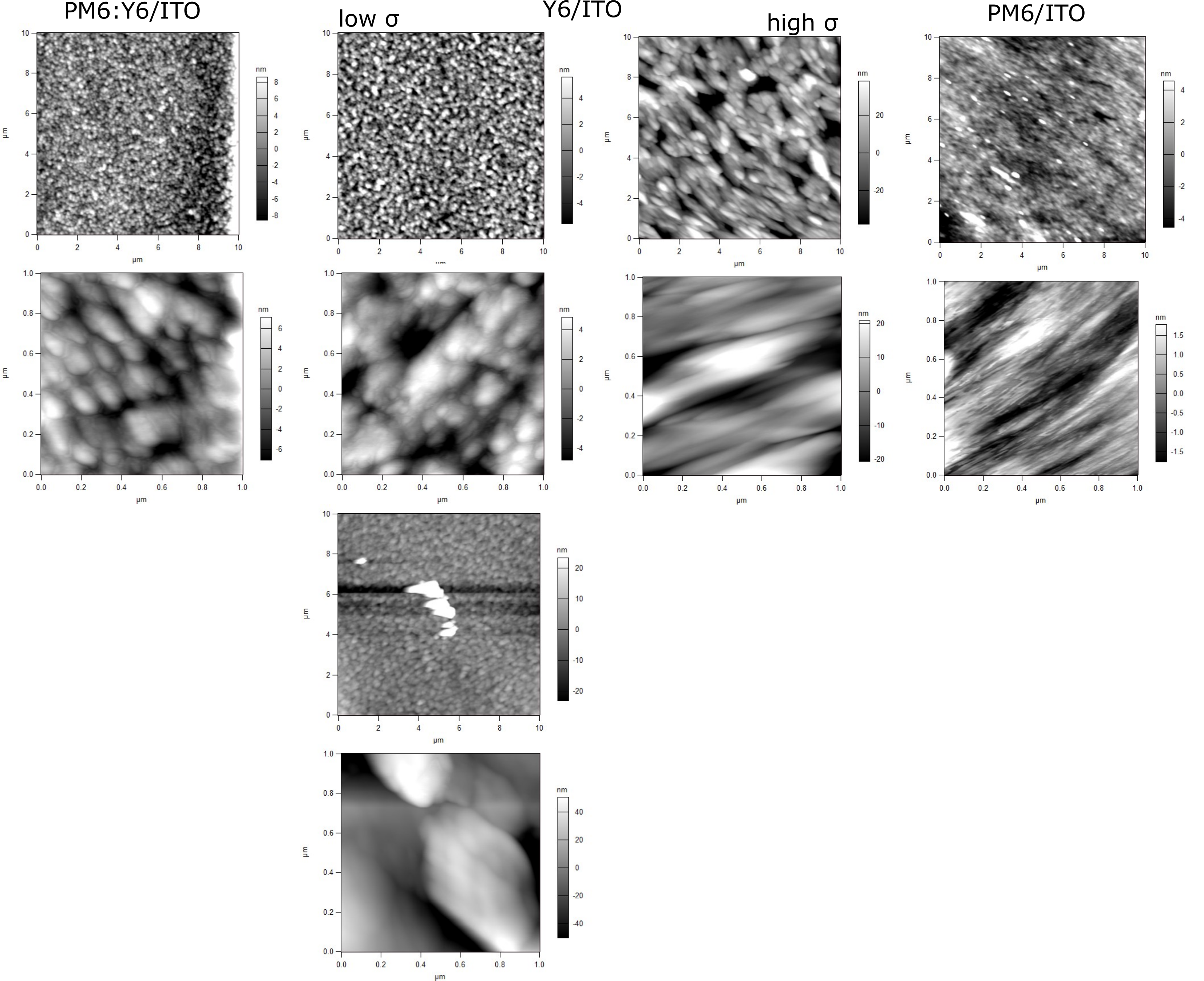}
    \end{adjustwidth}
    \caption[AFMs of PM6:Y6, Y6, and PM6 samples on ITO.]{AFMs of PM6:Y6, Y6, and PM6 samples on ITO. Y6/ITO ``low $\sigma$" is the lowest conductivity sample ($\square$, collected on two spots on the sample), and ``high $\sigma$" is one of the higher conductivity samples ($\triangle$).}
    \label{fig:AFMs-ito}
\end{figure}

Atomic force micrographs (AFMs) of polymer films were collected in air on a commercial instrument in tapping mode (Asylum Research MFP-3D-BIO) using a Olympus AC160TS-R3 probe. The AFMs (Figs.~\ref{fig:AFMs-pedot}--\ref{fig:AFMs-ito}) of the PEDOT:PSS/ITO samples give some insight into the conductivity. We observe that samples with higher conductivity have a lower rms roughness. This is noteable in the PM6:Y6/PEDOT:PSS/ITO samples, where the sample with lower conductivity ($\triangle$) has a roughness of 30 nm, and the sample with higher conductivity ($\square$) has a roughness of 6 nm.

\clearpage
\bibliography{main}